\documentclass{IEEEtran}
\makeatletter

\usepackage[noadjust]{cite}

\usepackage{mathtools,amssymb}
	\newtheorem{thm}{Theorem}
	\newtheorem{lem}[thm]{Lemma}
	\newtheorem{prop}[thm]{Proposition}
	\DeclareMathOperator*\depth{depth}
	\DeclareMathOperator*\bC{bC}

\usepackage{tikz}
	\definecolor{encolor}{HTML}{BBEECC}
	\definecolor{decolor}{HTML}{DEC0DE}
	\tikzset{
		every picture/.style={
			cap=round,join=round,
			baseline={([yshift=-.5ex]current bounding box)},
		},
		rp/.style={remember picture},o/.style={overlay},
		hl/.style={help lines},
		<^/.style={above left},^/.style={above},^>/.style={above right},
		<</.style={      left}                 ,>>/.style={      right},
		<_/.style={below left},_/.style={below},_>/.style={below right},
	}
	\tikzset{
		bb/.style={/utils/exec=\bbsetupcs,nodes={inner ysep=1,inner xsep=0}},
		G/.pic={\def\coor{coordinate}\path
			(-  \bbwire,   \bbgape)\coor(1L)(   \bbwire,   \bbgape)\coor(1R)
			(-  \bbwire,-  \bbgape)\coor(0L)(   \bbwire,-  \bbgape)\coor(0R);},
		F/.pic={\draw[pic actions]pic{G}(1L)--(1R)(0L)--(0R);},
		E/.pic={\pic[thick]{F};\draw[thick,fill=encolor]
			(-  \bbsize,-  \bbsize)rectangle(   \bbsize,   \bbsize)
			(-  \bbgape,-  \bbgape)       --(   \bbgape,   \bbgape);},
		D/.pic={\pic[thick]{F};\draw[thick,fill=decolor]
			(-  \bbsize,-  \bbsize)rectangle(   \bbsize,   \bbsize)
			(-  \bbgape,   \bbgape)       --(   \bbgape,-  \bbgape)..controls
			( .5\bbgape,-.5\bbgape)    and  ( .5\bbgape, .5\bbgape)..
			(   \bbgape,   \bbgape)       --(-  \bbgape,-  \bbgape);},
		inline/.style={
			every pic/.style={scale=5/6},
		}
	}
	\newdimen\bbsize\bbsize8pt  
	\newdimen\bbwire\bbwire12pt 
	\newdimen\bbgape\bbgape4pt  
	\pgfmathdeclarefunction{xoffset}{1}{\pgfmathparse{%
		((2^#1-1)*\bbwire-2*#1\bbgape)/3}}
	\def\bbsetupcs{
		\xdef\bbxmax{0}
		\def\pgfpointxyz##1##2##3{
			\ifnum##2>\bbxmax\xdef\bbxmax{##2}\fi
			\pgfpoint
				{(2\bbwire*##2+xoffset(\bbxmax)-xoffset(\bbxmax+1-##2))*##11}
				{0b##3*2\bbwire}
		}
	}
	\tikzset{
		ut/.style={
			/utils/exec=\utsetupcs,
			nodes={inner sep=1,minimum size=2,scale=5^(\utdepth)/6^(\utdepth)},
			baseline=.5\utupper-.5ex
		}
	}
	\def\utdepth{0}
	\newdimen\utupper\utupper8\bbwire
	\newdimen\utmiddle
	\newdimen\utlower\utlower0cm
	\def\utnode$#1$#2$#3#4{
		\utmiddle\dimexpr.5\utupper+.5\utlower\relax
		\draw(\utdepth,\utmiddle)node(\utdepth){$#1$}
			\ifnum\utdepth>0(\the\numexpr\utdepth-1)--(\utdepth)\fi;
		\edef\utdepth{\the\numexpr\utdepth+1}
		{\utlower\utmiddle#3}{\utupper\utmiddle#4}
		\draw[o](\utdepth-.333,\utmiddle)node{$#2$};
	}
	\def\utsetupcs{
		\def\pgfpointxy##1##2{\pgfpoint{2\bbwire*(1.2^(##1)*5-5)}{0pt}}
	}
	\def\TArule#1{\tikz\node[rounded corners,draw,align=center]{#1};}

\usepackage{pgfplots}
	\pgfplotsset{compat=1.14}
	\usepgfplotslibrary{units}
	\pgfplotsset{
		simu/.style={ymin=0,xmin=0}
	}

\usepackage[utf8]{inputenc}\let\DUC\DeclareUnicodeCharacter
	\DUC{03B1}\alpha     
	\DUC{03B2}\beta      
	\DUC{03B3}\gamma     
	\DUC{03B4}\delta     
	\DUC{03F5}\epsilon   
	\DUC{03B6}\zeta      
	\DUC{03B7}\eta       
	\DUC{03B8}\theta     
	\DUC{03B9}\iota      
	\DUC{03BA}\kappa     
	\DUC{03BB}\lambda    
	\DUC{03BC}\mu        
	\DUC{03BD}\nu        
	\DUC{03BE}\xi        
	\DUC{03C0}\pi        
	\DUC{03C1}\rho       
	\DUC{03C3}\sigma     
	\DUC{03C4}\tau       
	\DUC{03C5}\upsilon   
	\DUC{03D5}\phi       
	\DUC{03C7}\chi       
	\DUC{03C8}\psi       
	\DUC{03C9}\omega     
	\DUC{0393}\Gamma     
	\DUC{0394}\Delta     
	\DUC{03B5}\varepsilon
	\DUC{0398}\Theta     
	\DUC{03D1}\vartheta  
	\DUC{039B}\Lambda    
	\DUC{039E}\Xi        
	\DUC{03A0}\Pi        
	\DUC{03D6}\varpi     
	\DUC{03F1}\varrho    
	\DUC{03A3}\Sigma     
	\DUC{03C2}\varsigma  
	\DUC{03A5}\Upsilon   
	\DUC{03A6}\Phi       
	\DUC{03C6}\varphi    
	\DUC{03A8}\Psi       
	\DUC{03A9}\Omega     
	
	\DUC{1D53C}{\mathbb E}  
	\DUC{1D540}{\mathbb I}  
	\DUC{2119}{\mathbb P}   
	\DUC{1D49C}{\mathcal A} 
	\DUC{1D4AF}{\mathcal T} 
	
	\DUC{00B7}\cdot           
	\DUC{02C6}\hat            
	\DUC{2022}\bullet         
	\DUC{2190}\leftarrow      
	\DUC{2192}\to             
	\DUC{2208}\in             
	\DUC{2211}\sum            
	\DUC{221A}\sqrt           
	\DUC{221E}\infty          
	\DUC{2227}\wedge          
	\DUC{2228}\vee            
	\DUC{222A}\cup            
	\DUC{2248}\approx         
	\DUC{2254}\coloneqq       
	\DUC{2260}\neq            
	\DUC{2264}\le             
	\DUC{2265}\ge             
	\DUC{2282}\subset         
	\DUC{266D}{^-}            
	\DUC{266F}{^+}            
	\DUC{27F5}\longleftarrow  
	\DUC{27F6}\longrightarrow 
	
	\def\({\bigl(}
	\def\){\bigr)}
	\DUC{FF08}{\Bigl(}   
	\DUC{FF09}{\Bigr)}   
	\DUC{FF5B}{\Bigl\{}  
	\DUC{FF5D}{\Bigr\}}  
	\DUC{300C}{\biggl\{} 
	\DUC{300D}{\biggr\}} 

\usepackage[pdfusetitle,pdfsubject={Information Theory (cs.IT)}]{hyperref}

	\def\G(#1){pic(#1){G}}
	\def\F(#1){pic(#1){F}}
	\def\E(#1){pic(#1){E}}
	\def\D(#1){pic(#1){D}}
	\def\latin{\emph}
	\renewcommand\[{\@ifstar{\begin{equation*}}{\begin{equation}}}
	\renewcommand\]{\@ifstar{\end  {equation*}}{\end  {equation}}}
	\def\Capacity{\textnormal{Capacity}}
	\def\Bha{Bhattacharyya}
	\def\Arikan{Ar\i kan}
	\def\Ari{\text{Ar\i}}
	\def\T{T_\Ari}
	\def\loglogε{\log\lvert\log\varepsilon\rvert}
	\def\llol{[\mkern1mu{}^1_1{}^0_1]}
	\def\constant{\text{constant}}

\begin{document}
	\title {Log-logarithmic Time Pruned Polar Coding}
	\author{%
		Hsin-Po Wang and Iwan Duursma%
		\thanks{ University of Illinois at Urbana--Champaign}%
		\thanks{ \{hpwang2, duursma\} @illinois.edu}%
	}

\maketitle

\begin{abstract}
	A pruned variant of polar coding is proposed for binary erasure channels.
	For sufficiently small $ε>0$,
	we construct a series of capacity achieving codes with
	block length $N=ε^{-5}$, code rate $R=\Capacity-ε$, error probability $P=ε$,
	and encoding and decoding time complexity $\bC=O(\loglogε)$
	per information bit (Theorem~\ref{thm:main}).
	
	The given per-bit complexity~$\bC$ is log-logarithmic
	in $N$, in $\Capacity-R$, and in $P$;
	no known family of codes possesses this property.
	It is also the second lowest $\bC$
	after repeat-accumulate codes and their variants.
	While random codes and classical polar codes
	are the only two families of capacity-achieving codes
	whose $N$, $R$, $P$, and $\bC$ were written down as explicit functions,
	our construction gives the third family.
	
	Then we generalize the result to: Fix a prime $q$
	and fix a $q$-ary-input discrete symmetric memoryless channel.
	For sufficiently small $ε>0$,
	we construct a series of capacity achieving codes with
	block length $N=ε^{-O(1)}$,
	code rate $R=\Capacity-ε$,
	error probability $P=ε$,
	and encoding and decoding time complexity
	$\bC=O(\loglogε)$ per information bit
	(Theorem~\ref{thm:qmain}).
	The later construction gives the fastest family
	of capacity-achieving codes to date on those channels.
\end{abstract}

{
\catcode`\ =12
\def\_{\noexpand\_}\def\ {\noexpand\ }
	\message{^^J%
		       __                __           _______                   ^^J%
		      / /   ____  ____ _/ /___  ____ /_  __(_)___ ___  ___      ^^J%
		     / /   / __ \/ __ `/ / __ \/ __ `// / / / __ `__ \/ _ \     ^^J%
		    / /___/ /_/ / /_/ / / /_/ / /_/ // / / / / / / / /  __/     ^^J%
		   /_____/\____/\__, /_/\____/\__, //_/ /_/_/ /_/ /_/\___/      ^^J%
		               /____/        /____/                             ^^J%
	}
}

\section{Introduction}

	\IEEEPARstart{I}{n the}
	theory of two-terminal error correcting codes,
	four of the most essential parameters of block codes are
	block length~$N$, code rate~$R$, error probability~$P$,
	and per-bit time complexity~$\bC$.
	We brief the history below followed by our contribution over existing works.
	
	On day one, Shannon proved that for any communication channel,
	there exists a series of block codes such that
	$R$ approaches a number denoted by $\Capacity$ and $P$ converges to $0$.
	This property is called \emph{capacity achieving}.
	The price of achieving capacity is that $N$ must approach infinity,
	i.e., it is not possible to achieve capacity at finite block length.
	Another price is that $\bC$ grows exponentially in $N$
	by the nature of random coding.
	This makes Shannon's (and Fano and Gallager's) construction
	unsuitable for practical purposes.
	
	Coding theorists characterize how fast does the triple $(N,R,P)$
	approach $(∞,\Capacity,0)$, extending Shannon's theory.
	They treat $R(N),P(N)$ as functions in $N$
	and argue about the asymptote of both functions.
	They showed that $P$ alone can be as good as $2^{-N}$
	(\emph{error exponent regime}).
	They also showed that $R$ alone can be as good as $\Capacity-N^{-1/2}$
	(\emph{scaling exponent regime}).
	But together it is impossible to achieve
	$(R,P)=(\Capacity-N^{-1/2},2^{-N})$ at once.
	The correct asymptote is
	$(R,P)=（\Capacity-N^{-\constant},2^{-N^{\constant}}）$.
	This later paradigm is called \emph{moderate deviations regime}
	borrowed from probability theory.
	All three regimes inherit random coding as the main tool from Shannon,
	so $\bC$ is still on the order of $2^N$.
	See \cite{AW10,PV10,AW14,Arikan15,HT15} for recent progress.
	
	Beyond random coding, Reed--Muller code
	is one of the earliest codes with explicit construction.
	Its $N,R$ are easy to characterize.
	Beyond $N,R$, various decoding algorithms are proposed,
	each giving its own trade-off among $N,R,P,\bC$.
	Among them the most significant one is that Reed--Muller codes
	achieve capacity under MAP decoding over binary erasure channels (BEC)
	by Kudekar \latin{et al.}\ published in 2017 \cite{KKMPSU17}.
	That they achieve capacity is worthwhile by itself so the authors
	do not continue to write down the parametrization of $N,R,P$ explicitly.
	($\bC$ follows from Gaussian elimination.)
	That being said, we believe it is possible
	to infer the parametrization from their proof.
	
	On a different track, low density parity check (LDPC) codes are invented
	to generate codes with proper $(N,R,P,\bC)$-quadruples for practical use.
	The construction of LDPC codes gives the priority to $\bC$, so $\bC$ is low.
	But it is difficult to infer any parametrization of $N,R,P$.
	It was only recently, in 2013, that Kudekar \latin{et al.}\ 
	proved that LDPC codes achieve capacity \cite{KRU13}.
	Yet, their proof does not explicitly parametrize $N,P$.
	Even more extremely, a variant of LDPC codes called repeat-accumulate codes
	puts all efforts on reducing $\bC$.
	They finally arrived at capacity achieving codes
	with bounded $\bC$ over BEC \cite{PSU05,PS07}.
	Bounded complexity is the best possibility because
	the encoder should at least read in all inputs.
	But, again, their proofs do not explicitly parametrize $N,P$.
	
	In 2009, \Arikan{} observed the phenomenon of \emph{channel polarization}
	and proposed accordingly polar codes \cite{Arikan09}.
	Using Doob's martingale convergence theorem, \Arikan{} is able to show
	that polar codes achieve capacity with $\bC=O(\log N)$.
	Since then, researchers try to tune polar codes
	and characterize the corresponding $(N,R,P,\bC)$ asymptote.
	They find that $P$ is on the order of $2^{-N^{\constant}}$
	and that $\Capacity-R$ is on the order of $N^{-\constant}$
	\cite{GX13,MHU16,FT17,WD18m,BGS18,WD18l}.
	(Just like random codes except that the constants are off.)
	In particular, the following choice of constants is realizable
	by a series of polar codes (see Lemma~\ref{lem:md} for details):
	\begin{align}
		&{} (N,R,P,\bC) \notag\\
		&= （N,\Capacity-N^{-1/4},2^{-N^{1/24}},O(\log N)）. \label{eq:exptrade}
	\end{align}
	
	Our main contribution is to construct a pruned variant of polar codes
	and characterize its $(N,R,P,\bC)$ asymptote.
	More precisely, for an arbitrary BEC,
	Theorem~\ref{thm:main} provides a series of pruned polar codes with
	\begin{align*}
		&{} (N,R,P,\bC) \\
		&= （N,\Capacity-N^{-1/5},N^{-1/5},O(\log\log N)） \\
		&= （ε^{-5},\Capacity-ε,ε,O(\loglogε)）.
	\end{align*}
	Here $ε>0$ is an auxiliary parameter meant to be small.
	As $ε→0$ this asymptote is clearly capacity achieving.
	In contrast to Asymptote~(\ref{eq:exptrade}), our pruned polar codes
	loosen $P$ from $2^{-N^{1/24}}$ to $N^{-1/5}$
	but improve $\bC$ from $O(\log N)$ to $O(\log\log N)$.
	The lowered $\bC$ is now log-logarithmic
	in $N$, in $P$, and in $\Capacity-R$.
	This justifies the title.
	This is the first time polar codes are tuned
	to have $\bC$ as low as $O(\log\log N)$.
	This is also, we believe, the very next code
	on the leader board of low complexity codes
	after repeat-accumulate codes and their weaker variants
	(decoding complexity not bounded in $\Capacity-R$, say).
	In terms of the $(N,R,P,\bC)$ asymptote,
	we believe that this is the third time a family of block codes
	has a parametrization of $(N,R,P,\bC)$-quadruples,
	after random coding and classical polar coding.
	(Or the fourth, Reed--Muller codes being the third.)
	
	Here is a brief summary of the proof technique:
	we mentioned above that \Arikan{} observed
	the channel polarization phenomenon.
	The phenomenon is caused by the channel transformation $\T$.
	What $\T$ does is to transform a channel into two other channels,
	one of them has its \Bha{} parameter squared.
	After $n$ rounds of applying $\T$, the majority of good channels
	has gone through roughly $n/2$ times of squaring.
	Thus the \Bha{} parameters of these good channels
	are on the order of $2^{-2^{n/2}}$ \cite{AT09}.
	We realize that it takes only $O(\log n)$ times of squaring
	to achieve the order of $2^{-2n}$.
	An order of $2^{-2n}$ suffices for achieving capacity;
	the remaining applications of $\T$ can be pruned.
	Since on average we prune all but $O(\log n)$ many applications of $\T$,
	the per-bit time complexity is
	$\bC=O(\log n)=O(\log\log 2^n)=O(\log\log N)$.
	See Section~\ref{sec:main} for details.
	
	That $\T$ can be pruned is not our novel idea.
	Recent works on the implementation of polar coding develop
	a toolbox of engineering gadgets (including pruning) that
	accelerate the performance of polar codes in the real world.
	See Section~\ref{sec:connect} for what other researchers have done
	and how their ideas, when combined, can motivate our result.
	Alongside their huge success in optimization,
	we analyze the mathematical asymptote for the first time.
	Our result explains why pruning is inevitable and powerful,
	pointing out a new direction to faster (polar) codes.
	
	Last but not the least, as polar coding is generalized to other channels,
	we generalize our result to BSC, B-DMC, and more non-binary channels
	in Theorem~\ref{thm:qmain}.
	From our point of view, this is a very hard work since
	the preliminary result we need for BEC does not even have a BSC counterpart.
	(We end up proving them barehanded.)
	For readers not keen on details,
	it suffices to know that the introduced log-logarithm complexity
	is not unique to BECs.
	It is a rather universal phenomenon that channels polarize---%
	all but polynomially many of them polarize doubly-exponentially fast---%
	and pruning $\T$ is a universal technique that harvests
	channels as early as when they are sufficiently polarized.
	When done properly, pruning ends up with the fastest family
	of capacity-achieving codes on general channels.
	
	Organization:
	Section~\ref{sec:prelim} reviews channel polarization
	and introduces a general tree notation for later use.
	Section~\ref{sec:main} develops the main result, Theorem~\ref{thm:main}.
	Section~\ref{sec:connect} connects our work with others'.
	Section~\ref{sec:qchannel} extends the result to
	$q$-ary-input discrete symmetric memoryless channels
	for any prime $q$, concluding at Theorem~\ref{thm:qmain}.

\section{Preliminary} \label{sec:prelim}

\subsection{Channel Polarization and Tree Notation}

	Channel polarization \cite{Arikan09} is a method to synthesize some channels
	to form some extremely-reliable channels and some extremely-risky channels.
	The user then can transmit uncoded messages through extremely-reliable ones
	while padding predictable symbols through extremely-risky ones.
	We summarize channel polarization as follows.
	
\begin{figure}
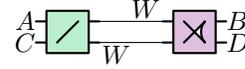

		$$\tikz[bb]\draw
			(-,1,0)\E(E)(+,1,0)\D(D)
			(E1R)--node[^>]{$W$}(D1L)
			(E0R)--node[<_]{$W$}(D0L)
			(E1L)node[<<]{$A$}(D1R)node[>>]{$B$}
			(E0L)node[<<]{$C$}(D0R)node[>>]{$D$}
		;$$
		\caption{
			The starting point of polar code construction.
			Two horizontal lines marked $W$
			are two independent copies of the BEC $W$.
			Pin $A$ to pin $B$ form a BEC which is denoted by $W♭$.
			It is a synthetic channel that is more risky than $W$.
			Pin $C$ to pin $D$ form another BEC which is denoted by $W♯$.
			It is a synthetic channel that is more reliable than $W$.
			Cf.\ \cite[Fig.~1]{Arikan09}.
		}
		\label{fig:cir2}
	\end{figure}

	Say we are going to communicate over a BEC $W$.
	One of \Arikan's contributions is the abstraction of two
	\emph{butterfly devices} \tikz[inline]\pic{E}; and \tikz[inline]\pic{D};.
	(Cf.\ \cite[Fig. 9, 10, and~5]{Arikan09}.)
	The butterfly devices work in a way that
	when we wire two independent copies of $W$ like Fig.~\ref{fig:cir2} does,
	pin $A$ and $B$ form a more risky synthetic channel $W♭$
	while pin $C$ and $D$ form a more reliable synthetic channel $W♯$.
	
\begin{figure}
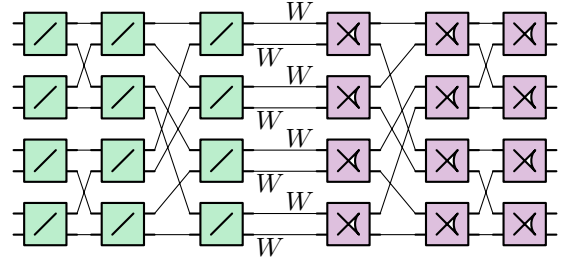

		$$\tikz[bb]{\draw
			foreach\J in{00,01,10,11}{
				foreach\l in{3,2,1}{
					(-,\l,\J)\E(\l E\J)(+,\l,\J)\D(\l D\J)}
				(1E\J1R)--node[^>]{$W$}(1D\J1L)
				(1E\J0R)--node[<_]{$W$}(1D\J0L)};
			\def\DRAW#1#2{
				\draw foreach\J in{000,001,010,011,100,101,110,111}{
					(#2E\K R)--(#1E\J L)(#1D\J R)--(#2D\K L)};}
			\def\K{\expandafter\k\J}
			\def\k#1#2#3{#3#1#2}\DRAW12
			\def\k#1#2#3{#1#3#2}\DRAW23
		}$$
		\caption{
			Fig.~\ref{fig:cir2} works like a recursive function.
			We can call the function three times to obtain this circuit.
			At the middle column we have eight independent copies of BEC $W$.
			The inner layer of butterfly devices will turn them into
			four independents copies of $W♭$ and
			four independents copies of $W♯$.
			The second layer of butterfly devices will turn them into
			$(W♭)♭$, $(W♭)♯$, $(W♯)♭$, and $(W♯)♯$,
			each of two independent copies.
			Finally the outer layer of butterfly devices will turn them into
			$((W♭)♭)♭$, $((W♭)♭)♯$, $((W♭)♯)♭$, $((W♭)♯)♯$,
			$((W♯)♭)♭$, $((W♯)♭)♯$, $((W♯)♯)♭$, and $((W♯)♯)♯$.
			Cf.\ \cite[Fig. 2 and~3]{Arikan09}.
		}
		\label{fig:cir8}
	\end{figure}

	\Arikan{} treats Fig.~\ref{fig:cir2} as a recursive function
	where nested calls to the function will
	generate circuits like Fig.~\ref{fig:cir8}.
	In particular, the circuit in Fig.~\ref{fig:cir8}
	generates eight synthetic channels
	$((W♭)♭)♭$, $((W♭)♭)♯$, and all the way up to $((W♯)♯)♯$.
	As the circuit gets larger and larger,
	we will end up getting $2^{\text{number of calls}}$ channels,
	from $(\dotso(W♭)♭\dotso)♭$ to $(\dotso(W♯)♯\dotso)♯$.
	\Arikan{} observes that synthetic channels generated in this way
	tend to be either extremely reliable or extremely risky.
	That is, they \emph{polarize}.
	He calls this phenomenon \emph{channel polarization}.
	
\begin{figure}
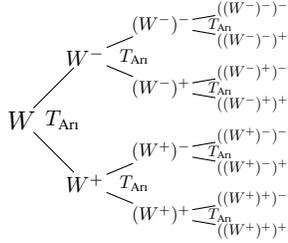

		$$\tikz[ut]{
			\utnode$W$\T${
				\utnode$W♭$\T${
					\utnode$(W♭)♭$\T${
						\utnode$((W♭)♭)♭$${}{} }{
						\utnode$((W♭)♭)♯$${}{} } }{
					\utnode$(W♭)♯$\T${
						\utnode$((W♭)♯)♭$${}{} }{
						\utnode$((W♭)♯)♯$${}{} } } }{
				\utnode$W♯$\T${
					\utnode$(W♯)♭$\T${
						\utnode$((W♯)♭)♭$${}{} }{
						\utnode$((W♯)♭)♯$${}{} } }{
					\utnode$(W♯)♯$\T${
						\utnode$((W♯)♯)♭$${}{} }{
						\utnode$((W♯)♯)♯$${}{} } } }
		}$$
		\caption{
			In Fig.~\ref{fig:cir8} we explain
			how the circuit transforms a channel to another.
			This operation can be encoded by a tree with auxiliary labels.
			In the tree, each vertex is a channel.
			A vertex is either a leaf or has two children.
			When a channel has two children,
			they form a parent-child-child triangle
			which represents the fact that the parent channel, say $w$,
			is transformed into $w♭$ (upper child) plus $w♯$ (lower child)
			by the butterfly devices.
			Instead of verbosely spamming ``butterfly devices,''
			we put a $\T$ at the center of each such triangle.
			It represents that
			butterfly devices serve as a channel transformation and that
			it is \Arikan{} who first recognizes/invents this transformation.
		}
		\label{fig:tre8}
	\end{figure}

	The relation among $W,W♭,\dotsc,((W♯)♯)♯$
	is summarized by a channel transformation $\T$
	as is discussed in \cite[Section~II]{Arikan09}.
	We reproduce and improve \cite[Fig.~6]{Arikan09} in Fig.~\ref{fig:tre8}.
	It is a tree whose vertexes are channels.
	Each parent-child-child triple represents the fact that
	the butterfly devices turn two independent copies of the parent channel
	into an upper child channel and a lower child channel.
	
	We introduce in the next subsection that it is possible
	to prune circuits and trees to reduce complexity.
	We will take advantage of the fact that circuits and trees
	correspond to each other and only argue about trees.
	Eventually we will show how we skillfully prune trees
	without having to sacrifice $R$ and $P$ too much.

\subsection{Pruning Circuits and Trees}

	The observation that circuits and trees can be pruned
	to attain a different $(R,P,\bC)$ trade-off of codes
	has been made several times in the past.
	For instance, Fig.~\ref{fig:prune7} illustrates a circuit-tree pair
	that saves two butterfly devices, which potentially saves some time.
	Fig.~\ref{fig:prune5} illustrates another circuit-tree pair
	that saves six butterfly devices, which potentially saves more time.
	
	\begin{figure}
		$$\tikz[bb]{
			\foreach\J in{00,01,10,11}{
				\draw foreach\l in{3,2,1}{\ifnum\J\l=113
						(-,\l,\J)\F(\l E\J)(+,\l,\J)\F(\l D\J)\else
						(-,\l,\J)\E(\l E\J)(+,\l,\J)\D(\l D\J)\fi}
					(1E\J1R)--node[^>]{$W$}(1D\J1L)
					(1E\J0R)--node[<_]{$W$}(1D\J0L);}
			\def\DRAW#1#2{
				\draw foreach\J in{000,001,010,011,100,101,110,111}{
					(#2E\K R)--(#1E\J L)(#1D\J R)--(#2D\K L)};}
			\def\K{\expandafter\k\J}
			\def\k#1#2#3{#3#1#2}\DRAW12
			\def\k#1#2#3{#1#3#2}\DRAW23
		}$$$$\tikz[ut]{
			\utnode$W$\T${
				\utnode$W♭$\T${
					\utnode$(W♭)♭$${}{} }{
					\utnode$(W♭)♯$\T${
						\utnode$((W♭)♯)♭$${}{} }{
						\utnode$((W♭)♯)♯$${}{} } } }{
				\utnode$W♯$\T${
					\utnode$(W♯)♭$\T${
						\utnode$((W♯)♭)♭$${}{} }{
						\utnode$((W♯)♭)♯$${}{} } }{
					\utnode$(W♯)♯$\T${
						\utnode$((W♯)♯)♭$${}{} }{
						\utnode$((W♯)♯)♯$${}{} } } }
		}$$
		\caption{
			The top part is a pruned circuit where
			the butterfly devices applied to $(W♭)♭$ are dropped.
			As a result, this circuit does not generate $((W♭)♭)♭$ or $((W♭)♭)♯$
			and leaves the two copies of $(W♭)♭$ intact.
			The complete list of generated channels reads:
			$(W♭)♭$, $(W♭)♭$, $((W♭)♯)♭$, $((W♭)♯)♯$,
			$((W♯)♭)♭$, $((W♯)♭)♯$, $((W♯)♯)♭$, $((W♯)♯)♯$.
			The bottom part is a pruned tree that illustrates the fact that
			$(W♭)♭$ does not undergo the third round of application of $\T$
			and has no children.
			On the other hand, other ``depth-$2$'' channels
			$(W♭)♯$, $(W♯)♭$, $(W♯)♯$ undergo $\T$ and
			generate what they used to generate in Fig.~\ref{fig:tre8}.
		}
		\label{fig:prune7}
	\end{figure}
	
	\begin{figure}
		$$\tikz[bb]{
			\foreach\J in{00,01,10,11}{
				\draw foreach\l in{3,2,1}{
					\ifnum\l=3
						\ifnum\J=10
							(-,\l,\J)\E(\l E\J)(+,\l,\J)\D(\l D\J)
						\else
							(-,\l,\J)\F(\l E\J)(+,\l,\J)\F(\l D\J)
						\fi
					\else
						(-,\l,\J)\E(\l E\J)(+,\l,\J)\D(\l D\J)
					\fi}
					(1E\J1R)--node[^>]{$W$}(1D\J1L)
					(1E\J0R)--node[<_]{$W$}(1D\J0L);}
			\def\DRAW#1#2{
				\draw foreach\J in{000,001,010,011,100,101,110,111}{
					(#2E\K R)--(#1E\J L)(#1D\J R)--(#2D\K L)};}
			\def\K{\expandafter\k\J}
			\def\k#1#2#3{#3#1#2}\DRAW12
			\def\k#1#2#3{#1#3#2}\DRAW23
		}$$$$\tikz[ut]{
			\utnode$W$\T${
				\utnode$W♭$\T${
					\utnode$(W♭)♭$${}{} }{
					\utnode$(W♭)♯$\T${
						\utnode$((W♭)♯)♭$${}{} }{
						\utnode$((W♭)♯)♯$${}{} } } }{
				\utnode$W♯$\T${
					\utnode$(W♯)♭$${}{} }{
					\utnode$(W♯)♯$${}{} } }
		}$$
		\caption{
			The top part is a pruned circuit where the butterfly devices
			applied to $(W♭)♭$, $(W♯)♭$, and $(W♯)♯$ are dropped.
			They (each of two copies) are left intact.
			The bottom part is a pruned tree that encodes
			what happens in the circuit: only $(W♭)♯$
			undergoes the third round of application of $\T$ and has children.
			The complete list of generated channels reads:
			$(W♭)♭$, $(W♭)♭$, $((W♭)♯)♭$, $((W♭)♯)♯$,
			$(W♯)♭$, $(W♯)♭$, $(W♯)♯$, $(W♯)♯$.
		}
		\label{fig:prune5}
	\end{figure}
	
	Roughly speaking, we expect that
	the more we prune the circuit and tree, the more butterfly devices we save.
	This potentially saves even more time.
	However, the saving in time, if any,%
	\footnote{
		We keep emphasising ``potentially'' because an asymmetric design
		of encoder and decoder is more difficult to implement.
		It is totally possible that we end up not saving any wall-clock time
		because the asymmetric implementation is
		slower or too expensive to optimize.
	}
	does not come for free.
	Since the resulting synthetic channels
	are different form before, $P$ varies.
	Thus we have to recompute/remeasure $P$
	and then check whether we can resist the new $P$.
	One degenerate case is that we simply drop all butterfly devices;
	this saves 100\% of time but then there is no coding at all.
	
	See Section~\ref{sec:connect} for a list of former works which show that
	pruning empirically speeds up the encoding and decoding
	but does not sacrifice other parameters too much.
	Among them it is common to see claims of
	their architecture saving 50\% or 90\% of time, experimentally.
	Our contribution over existing works is that we quantize
	the trade-off mathematically instead of testing and measuring.
	As we will show that $\bC$ can be reduced
	from $O(\log N)$ to $O(\log\log N)$, this is a 99.9...\% save%
	\footnote{
		We are aware of the fact that the 5G standard,
		considered as the main application of polar codes,
		has a latency restriction.
		Thus $N$ is capped.
		Our results apply in a different---asymptotic---range .
	}
	as $N→∞$.
	
	In the next subsection we review the \Bha{} parameter $Z$
	and the processes $K_i,Z_i,I_i$, and then we generalize them.
	We will show how they relate to trees, especially to pruned trees.
	Being able to relate trees to processes makes it possible
	to control the behavior of codes properly.

\subsection{\Bha{} Parameter and Processes}

	The \emph{\Bha{} parameter} $Z(W)$ of a channel $W$
	measures the risk (unreliability) of the channel.
	For BEC, $Z(W)$ coincides with the erasure probability of $W$.
	The symmetric capacity $I(W)$ of $W$ coincides with the complement $1-Z(W)$.
	Recall the processes $K_i$, $I_i$, and $Z_i$ as defined in
	\cite[Section~IV, third paragraph]{Arikan09}.
	Therein $K_i$ is the process starting from $K_0≔W$; and
	$K_{i+1}$ is either $K_i♭$ or $K_i♯$, each with $1/2$ probability.
	The process of \Bha{} parameter $Z_i$ is $Z(K_i)$.
	The process of capacity $I_i$ is $I(K_i)$.
	Clearly $I_i+Z_i=1$; we are on BEC.
	Here is our generalization.
	
	Denote by $𝒯$ a finite rooted tree of channels with root channel $W$.
	We stick to the convention that: the root has depth $0$;
	the depth of a tree is the depth of the deepest leaf;
	and the tree with only one vertex has depth $0$.
	Therefore, the circuit corresponding to $𝒯$
	consumes $2^{\depth(𝒯)}$ root channels;
	and for any leaf channel $w$,
	the circuit generates $2^{\depth(𝒯)-\depth(w)}$ copies of $w$.
	
	Given a finite channel tree $𝒯$ with root channel $W$,
	define three discrete-time stochastic processes
	$K_{i∧τ}$, $Z_{i∧τ}$, $I_{i∧τ}$ and a stopping time $τ$ as follows:
	Start from the root channel $K_{0∧τ}≔W$.
	For any $i≥0$, if $K_{i∧τ}$ is a leaf, let $K_{i+1∧τ}$ be $K_{i∧τ}$.
	If, otherwise, $K_{i∧τ}$ has two children,
	choose either child with equal probability as $K_{i+1∧τ}$.
	Since we work on finite trees, there is always a smallest index $j$
	such that $K_{j∧τ}=K_{j+1∧τ}=K_{j+2∧τ}=$ \latin{ad infinitum}.
	Let $τ$ be this smallest index.
	Then $τ$ is the \emph{stopping time} that
	records when $K_{i∧τ}$ ``stops evolving.''
	Let $K_τ$ be the channel $K_{i∧τ}$ when it stops evolving.
	That is, $K_τ=K_{τ∧τ}=\lim_{i→∞}K_{i∧τ}$.
	Let $Z_{i∧τ}$ be $Z(K_{i∧τ})$.
	Let $I_{i∧τ}$ be $I(K_{i∧τ})$.
	Let $Z_τ$ be $Z(K_τ)=\lim_{i→∞}Z_{i∧τ}$.
	Let $I_τ$ be $I(K_τ)=\lim_{i→∞}I_{i∧τ}$.
	
	Readers familiar with probability theory will notice that
	the notation $K_{i∧τ},Z_{i∧τ},I_{i∧τ}$ coincide with what Gallager calls
	\emph{stopped process} \cite[Theorem 9.7.1]{Gallager13}.
	Other readers may stick to the operational definition presented above.
	
\begin{figure}
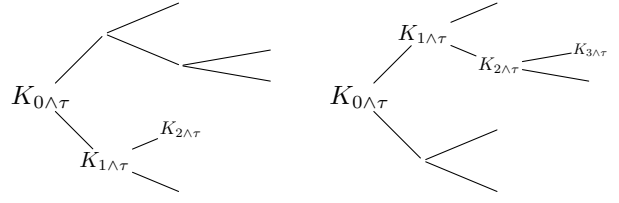

		$$\tikz[ut]{
			\utnode$K_{0∧τ}$${
				\utnode$$${
					\utnode$$${}{} }{
					\utnode$$${
						\utnode$$${}{} }{
						\utnode$$${}{} } } }{
				\utnode$K_{1∧τ}$${
					\utnode$K_{2∧τ}$${}{} }{
					\utnode$$${}{} } }
		}\qquad\tikz[ut]{
			\utnode$K_{0∧τ}$${
				\utnode$K_{1∧τ}$${
					\utnode$$${}{} }{
					\utnode$K_{2∧τ}$${
						\utnode$K_{3∧τ}$${}{} }{
						\utnode$$${}{} } } }{
				\utnode$$${
					\utnode$$${}{} }{
					\utnode$$${}{} } }
		}$$
		\caption{
			Recall the tree in Fig.~\ref{fig:prune5}.
			On the left is a possible trajectory of the process $K_{i∧τ}$.
			We begin with $K_{0∧τ}$ being the root channel $W$.
			It has children.
			The first ``coin toss'' chooses the lower child $W♯$ as $K_{1∧τ}$.
			It has children.
			The second coin toss chooses the upper child $(W♯)♭$ as $K_{2∧τ}$.
			It has no child.
			The process stabilizes.
			So $K_{2∧τ}=K_{3∧τ}=K_{4∧τ}=\dotsb=K_τ$ and $τ=2$.
			The probability measure of this trajectory is $1/8$.
			On the right is another possible trajectory
			of the process $K_{i∧τ}$.
			We begin with $K_{0∧τ}$ being the root channel.
			It has children.
			The first coin toss chooses the upper child $W♭$ as $K_{1∧τ}$.
			It has children.
			The second coin toss chooses the lower child $(W♭)♯$ as $K_{2∧τ}$.
			It has children
			The third coin toss chooses the upper child $((W♭)♯)♭$ as $K_{3∧τ}$.
			It has no child.
			The process stabilizes with
			$K_{3∧τ}=K_{4∧τ}=K_{5∧τ}=\dotsb=K_τ$ and $τ=3$.
			The probability measure of this trajectory is $1/4$.
		}
		\label{fig:traj}
	\end{figure}

	Recall the pruned tree in Fig.~\ref{fig:prune5}.
	We give two possible trajectories in Fig.~\ref{fig:traj}.
	Note that this tree is a nontrivial example where $τ$ is not a constant.
	As a random variable, $τ$ depends on
	which child of $K_{i∧τ}$ is chosen at each step.
	It turns out that $ℙ\{τ=2\}=3/4$ and $ℙ\{τ=3\}=1/4$.
	For the tree in Fig.~\ref{fig:prune7}, $ℙ\{τ=2\}=1/4$ and $ℙ\{τ=3\}=3/4$.
	For the tree in Fig.~\ref{fig:tre8}, however, $τ=3$ with probability $1$.
	
	By \cite[Proposition~8]{Arikan09}, $I_i$ is a martingale.
	Hence $I_{i∧τ}$ is a martingale by \cite[Theorem~5.2.6]{Durrett10}.
	Since $W$ is a BEC, $Z_{i∧τ}=1-I_{i∧τ}$ is a martingale as well.
	A useful consequence by applying
	\cite[Theorem~5.7.6]{Durrett10} to $I_i,1-I_i$ is
	\[I(W)=I_0=𝔼[I_τ]. \label{eq:conserv}\]
	Recall that $I_i$ being a martingale
	plays two crucial roles in \Arikan's proof.
	For one: the martingale convergence theorem applies.
	For two: $I(W)=I_0=𝔼[I_i]$ so $ℙ\{I_∞=1\}=I(W)$.
	Equation~(\ref{eq:conserv}) generalizes this argument in the manner that
	we can now decide whether to prune a branch or not
	on a channel-by-channel basis.
	This creates a new level of flexibility
	to balance $\bC$ and other parameters.
	
	In the next subsection we show how trees and processes relate to codes.
	Only after we establish the relation between trees and $(N,R,P,\bC)$
	can we optimize how we are going to prune the tree.

\subsection{From Trees to Codes and Communication}

	Recall that in a given tree $𝒯$, non-leaf vertexes represent
	channels that are consumed to obtain their children.
	They are not available to users.
	Leaves of $𝒯$, however, represent channels that are available to users.
	A user who wants to send messages using $𝒯$ can:
	1) choose a subset $𝒜$ of leaves;
	2) transmit uncoded messages through leaf channels in $𝒜$; and
	3) pad predictable symbols through the remaining leaf channels.
	
	This makes the tree-leaves pair $(𝒯,𝒜)$ a block code.
	We want to characterize this block code by analyzing these four parameters:
	block length~$N$, code rate~$R$, error probability~$P$, and time complexity.
	Here is how to read-off these parameters from $(𝒯,𝒜)$.
	
	The \emph{block length} $N$ of $(𝒯,𝒜)$ is
	the number of copies of $W$ in the circuit.
	In term of trees, it is
	\[*N≔2^{\depth(𝒯)}.\]*
	$N$ does not depend on $𝒜$, so we can talk about
	``the block length of $𝒯$'' without defining $𝒜$ in advance.
	
	The \emph{code rate} $R$ of $(𝒯,𝒜)$ is the number
	of synthetic channels in $𝒜$ (multiplicity included) divided by $N$.
	In terms of processes, it is the probability of $K_τ$ ending up in $𝒜$.
	\[*R≔ℙ\{K_τ∈𝒜\}.\]*
	
	The \emph{error probability} $P$ of $(𝒯,𝒜)$ is the probability
	that any leaf channel in $𝒜$ fails to transmit the message.
	For classical polar codes, error probability is at most
	$\sum_{w\in𝒜}Z(w)$ as stated in \cite[Proposition~2]{Arikan09}.
	For pruned polar codes, the error probability
	is at most a weighted sum as follows
	\[*P≤\sum_{w\in𝒜}Nℙ\{K_τ=w\}Z(w).\]*
	This is because $Z(w)$ bounds from above
	the error probability of the synthetic channel $w$.
	Thus it suffices to apply the union bound where
	$Z(w)$ is weighted by the multiplicity of $w$.
	Detailed proof omitted.
	
	The \emph{per-block time complexity} of $(𝒯,𝒜)$ is
	how long $𝒯$'s circuit takes to execute.
	It is bounded from above by the number of butterfly devices
	multiplied by the time each butterfly device spends.
	(No parallelism allowed.)
	The design of the butterfly devices suggests that
	each butterfly device spends constant time.
	Thus the per-block time complexity is
	proportional to the number of butterfly devices.
	As each leaf channel $K_τ$ at depth $τ$ must go through
	$τ$ \tikz[inline]\pic{E};'s and $τ$ \tikz[inline]\pic{D};'s,
	the total number of butterfly devices is exactly $2N𝔼[τ]$.
	Hence the per-block time complexity is proportional to $N𝔼[τ]$.
	
	The \emph{per-bit time complexity} $\bC$ is
	the amortized time each information bit should pay.
	Naturally it is proportional to $N𝔼[τ]/NR=𝔼[τ]/R$.
	In our case, we are persuading capacity achieving codes
	so $R≈I(W)$ is about a constant.
	Therefore, we infer that the per-bit time complexity is proportional to
	\[𝔼[τ]. \label{eq:Etau}\]
	$𝔼[τ]$ does not depends on $𝒜$, so we can talk about
	``the complexity of $𝒯$'' without defining $𝒜$ in advance.
	
	We are almost ready to show readers how to prune trees
	except that we will phrase pruning in a specific tone:
	Instead of starting from a huge, heavy tree
	and pruning 99.9...\% of its vertexes,
	we grow a tree from scratch and decide channel-by-channel
	whether each channel should have children or not.
	Doing so fits the stochastic processes paradigm more properly because
	usually we are not allowed to look into the future (see the descendants)
	before we make the decision (whether it should have children or not).
	We assure that this is a matter of wording style and
	has nothing to do with the actual properties of codes.
	
	In this context, we \emph{apply $\T$ to $w$}
	if we want $w$ to have children.
	We \emph{do not apply $\T$ to $w$}
	if we want the opposite, that $w$ should be a leaf.
	Here are two heuristic rules to keep in mind:
	1)	That $N≔2^{\depth(𝒯)}$ suggests that we should set a boundary $n$
		and do not apply $\T$ once we reach depth $n$.
		This guarantees that the block length $N$ will never exceed $2^n$.
		We assume the worst case scenario $N≔2^n$.
	2)	A mediocrely reliable channel increases $P$ too much if we utilize it,
		but sacrifices $R$ too much if we freeze it.
		Either way it becomes an obstacle to capacity achieving.
		To avoid the dilemma, the only chance is
		applying $\T$ to polarize them further.
		This suggests that we should make decision
		based on a threshold for ``mediocre reliability.''

\subsection{Growing Tree and Choosing Leaves as Code Construction}

	We showed how to estimate the parameters of a block code $(𝒯,𝒜)$
	if $𝒯$ and $𝒜$ are explicitly given.
	Now we state how we are going to grow a good tree of prescribed depth $n$
	(instead of pruning the perfect binary tree of depth $n$).
	Here $n$ is an integer to be assigned.
	Let $ε>0$ be small.
	Let $Y(w)$ be $\min\{Z(w),1-Z(w)\}$.
	
	Begin with $W$ as the only vertex of a new rooted tree.
	We announce the following framed rule:
	\[\TArule{
		Apply $\T$ to $w$ if and only if \\
		$\depth(w)<n$ and $Y(w)>ε2^{-n}$.
	} \label{eq:rule}\]
	The rule says: for each leaf $w$,
	if both $\depth(w)<n$ and $Y(w)>ε2^{-n}$ are met,
	apply $\T$ to $w$ to obtain $w♭$ and $w♯$;
	and then append $w♭$ and $w♯$ as children of $w$.
	If, otherwise, either criterion is not met,
	we do not apply $\T$ and leave $w$ as a leaf.
	See Appendix~\ref{app:grow} for a possible execution of the rule.
	We will see later that $Y(w)$ serves as a judgement
	of whether $w$ is sufficiently polarized or not.
	Having $𝒯$, we declare $𝒜$ by the criterion
	\[\TArule{
		$w∈𝒜$ if and only if \\
		$w$ is a leaf and $Z(w)≤ε2^{-n}$.
	} \label{eq:util}\]
	
	In Criterion~(\ref{eq:util}) and Framed Rule~(\ref{eq:rule}),
	we implicitly divide channels into three (actually four) classes:
	1)	For channels that are mediocrely reliable, i.e.,
		$ε2^{-n}<Z(w)<1-ε2^{-n}$, we apply $\T$ to polarize $w$ further.
	2)	For channels that are sufficiently reliable, i.e., $Z(w)≤ε2^{-n}$,
		we stop applying $\T$ and collect them in our pocket $𝒜$.
		Doing so as early as possible maximizes the save on butterfly devices.
		Nevertheless, every channel we put in $𝒜$
		contributes to the overall error probability $P$.
		We must choose wisely what to and what not to put in $𝒜$.
	3)	For channels that are incredibly risky, i.e. $1-ε2^{-n}≤Z(w)$,
		it becomes inefficient to extract the capacity from $w$.
		We should just ``let go'' the risky channels and save butterfly devices.
		The earlier we let it go the more butterfly devices we save.
		Nevertheless, since $𝔼[I_τ]$ is conservative,
		letting go a channel means giving up some capacity.
		We must not give up too much capacity as we want $R→I(w)$.
	4)	For channels that are mediocrely reliable \emph{at depth $n$},
		there is no chance to polarize them.
		We shall let it go.
	
	We now have both $𝒯$ and $𝒜$ properly defined.
	We will show in the coming section how $(𝒯,𝒜)$ performs.

\section{Main Result} \label{sec:main}

	We will prove the following.
	
	\begin{thm}[Main theorem] \label{thm:main}
		Assume any BEC $W$.
		For small enough $ε>0$,
		there exists a series of pruned polar codes with
		block length~$N$, code rate~$R$, error probability~$P$,
		and per-bit time complexity~$\bC$ satisfying
		\begin{align*}
			&{} (N,R,P,\bC) \\
			&= （N,I(W)-N^{-1/5},N^{-1/5},O(\log\log N)） \\
			&= （ε^{-5},I(W)-ε,ε,O(\loglogε)）.
		\end{align*}
	\end{thm}
	\begin{IEEEproof}
		The codes will be constructed in Theorem~\ref{thm:grow}.
		Proposition~\ref{prop:N} will compute its block length~$N$.
		Proposition~\ref{prop:C} will compute its per-bit time complexity~$\bC$.
		Proposition~\ref{prop:P} will compute its error probability~$P$.
		Proposition~\ref{prop:R} will compute its code rate~$R$.
		Together they certify the code satisfies the claimed asymptote.
	\end{IEEEproof}
	
	The general strategy is to grow a tree as Framed Rule~(\ref{eq:rule}) stated
	and choose leaves as Criterion~(\ref{eq:util}) stated.
	After that we control how $K_τ$ behaves.
	In order to control how $K_τ$ behaves,
	we need to learn how $K_i$ behaves.
	The following lemma is one of the early results
	that characterize how fast do synthetic channels polarize.
	It describes a phenomenon that ultimately leads to our result.
	
	\def\mag#1{\raisebox{-2pt}{\Large$\displaystyle#1$}}
	\begin{lem}\cite[Theorem~1]{GX13}
		There exists $μ'>0$ such that
		\[*ℙ「\mag{Z_i≤2^{-2^{0.49i}}}」≥I(W)-O(2^{-i/μ'}).\]*
	\end{lem}
	
	Intuitively speaking, this lemma shows that
	$Z_i$ goes to zero exponentially fast.
	Recall that in Framed Rule~(\ref{eq:rule})
	we do not apply $\T$ if $Z_i≤ε2^{-n}$.
	Here $ε2^{-n}$ is polynomial in $N$
	so $Z_i$ will reach this threshold in log-logarithmic steps.
	This is the main reason why the complexity is log-logarithmic.
	
	This lemma is later generalized to a form
	with explicit constants as follows.
	
	\begin{lem}\cite[Theorem~3 and Inequality~(56)]{MHU16}
		For $μ=3.627$ and $γ$ such that $1/(1+μ)<γ<1$,
		\[*ℙ「\mag{Z_i≤2^{-2^{iγh_2^{-1}(\frac{γμ+γ-1}{γμ})}}}」
			≥I(W)-O(2^{\frac{-i(1-γ)}{μ}}).\]*
	\end{lem}
	
	Here $h_2^{-1}$ is the inverse function of the binary entropy function;
	and $μ$ is a constant called \emph{scaling exponent}.
	\cite{FV14} gives the approximation $μ=3.627$.
	This lemma makes it possible to parametrize $N=ε^{-5}$
	instead of $N=ε^{-3μ}$ for some existing but unknown constant $μ$.
	We have to fall back to $N=ε^{-3μ}$ in the general channel case
	because such result does not exist.
	
	Although constants provided by the previous lemma
	suffices to prove an explicit-constant version of our theorem,
	we think it is useful to present an even stronger lemma.
	This lemma from our previous work gives stronger constants.
	We believe these constants are optimal.
	
	\begin{lem} \label{lem:md}
	\cite[Theorem~6]{WD18m}
		Fix $μ',β'$. If
		\[\frac{1-π}{μ'-μπ}+h_2（\frac{β'μ'}{μ'-μπ}）<1 \label{eq:picri}\]
		for all $π∈[0,1]$, then
		\[ℙ「\mag{Z_i≤2^{-2^{iβ'}}}」≥I(W)-O(2^{-i/μ'}). \label{eq:betamu}\]
	\end{lem}
	
	Here $h_2$ is the binary entropy function; and $μ=3.627$.
	We choose $(μ',β')=(4,1/24)$.
	Now Inequality~(\ref{eq:picri}) becomes
	\[\frac4{4-3.627π}+h_2（\frac{1/6}{4-3.627π}）<1. \label{eq:plug424}\]
	It holds for all $π∈[0,1]$;
	this is verified numerically in Appendix~\ref{app:numerical}.
	Thus Inequality~(\ref{eq:betamu}) becomes
	\[*ℙ｛Z_i≤2^{-2^{i/24}}｝≥I(W)-O(2^{-i/4}).\]*
	Since we are on BECs, the ``flipped version''
	\[*ℙ｛I_i≤2^{-2^{i/24}}｝≥Z(W)-O(2^{-i/4})\]*
	also holds.
	Together they capture the cases
	when $Z_i$ is (doubly exponentially) small and
	when $Z_i$ is (doubly exponentially) close to $1$.
	What is left is when $Z_i$ is mediocre.
	Let $Y_i$ be $\min\{Z_i,1-Z_i\}$, then
	\[ℙ｛Y_i>2^{-2^{i/24}}｝=O(2^{-i/4}). \label{eq:Y24O4}\]
	While $Z_i,I_i$ captures how (un)reliable channels are,
	$Y_i$ captures how far $Z_i$ and $I_i$ are away from their destination.
	In other words, $Y_i$ measures the extent of polarization.
	This makes $Y_i$ a more suitable variable
	for capturing the speed of convergence,
	while $Z_i,I_i$ serve the purpose of
	judging how (un)reliable a channel is.
	
	We are ready to analyze the stated construction.
	We will first prove a theorem regarding $𝔼[τ]$
	and then analyze $N,\bC,P,R$ in that order.
	Once we can control all four parameters
	we obtain the main theorem, Theorem~\ref{thm:main}.
	
	\begin{thm} \label{thm:grow}
		Given $W$ and $ε$.
		Assign $n≔-5\log_2ε$.
		Then Framed Rule~(\ref{eq:rule}), i.e.,
		\[*\TArule{
			Apply $\T$ to $w$ if and only if \\
			$\depth(w)<n$ and $Y(w)>ε2^{-n}$,
		}\]*
		grows a channel tree $𝒯$ with $𝔼[τ]=O(\loglogε)$.
	\end{thm}
	\begin{IEEEproof}
		Let us grow the tree and observe the processes $K_{i∧τ}$ ans $Z_{i∧τ}$.
		By the rule, channel $K_{i∧τ}$ has children
		if and only if $\depth(K_{i∧τ})<n$ and $Y(K_{i∧τ})>ε2^{-n}$.
		Conversely, channel $K_{i∧τ}$ has no child if and only if
		$\depth(K_{i∧τ})≥n$ or $Y(K_{i∧τ})≤ε2^{-n}$.
		The stopping time $τ$, by definition, is
		the least index $j$ such that $K_{j∧τ}$ has no child.
		So $τ$ is the least index $j$ such that
		$\depth(K_j)≥n$ or $Y(K_j)≤ε2^{-n}$.
		Equivalently, $τ$ is the least index $j$ such that
		$j≥n$ or $Y_j≤ε2^{-n}$.
		More formally,
		\[*τ=\min\(\{j:Y_j≤ε2^{-n}\}∪\{n\}\). \label{eq:stop}\]*
		
		For stopping times defined in the form
		``when is the first time something happens,''
		they are usually studied through the event $\{τ>i\}$.
		In other words, knowing ``when does something first happen''
		is equivalent to knowing ``whether something had happened before $i$.''
		In our case, the event $\{τ>i\}$ is equivalent to
		whether $i≥n$ or whether $Y_j≤ε2^{-n}$ for some $j≤i$.
		We just want an upper bound, so we check the largest index:
		whether $Y_i≤ε2^{-n}$ or not.
		More Formally
		\[*\{τ>i\}⊂\{Y_i>ε2^{-n}\}=\{Y_i>ε^6\}.\]*
		The equality is due to our choice of $n≔-5\log_2ε$.
		
		Whether $Y_i>ε^6$ or not can be relaxed to the disjunction
		$Y_i>2^{-2^{i/24}}$ or $2^{-2^{i/24}}>ε^6$.
		We have seen the first disjunct before, in Estimate~(\ref{eq:Y24O4}).
		The second disjunct is new, but we can solve for $i$ and
		deduce that $2^{-2^{i/24}}>ε^6$ implies that $i<O(\loglogε)$.
		More formally,
		\[*\{τ>i\}⊂｛Y_i>2^{-2^{i/24}}\text{ or }i<O(\loglogε)｝.\]*
		Now whether $τ>i$ happens or not is divided into two cases:
		1)	if $i$ is small enough such that $i<O(\loglogε)$,
			then we have little idea whether $τ>i$ or not (possibly not).
			This is not an accident though;
			we do not expect decent polarization at depth $O(\loglogε)$.
		2)	if $i$ is large enough to violate $i<O(\loglogε)$, then
			$\{τ>i\}$ is dominated by the first disjunct, $Y_i>2^{-2^{i/24}}$.
			Estimate~(\ref{eq:Y24O4}) bounds the probability measure from above.
		Put 1) and 2) together we have a joint bound
		\[*ℙ\{τ>i\}≤\begin{cases*}
			1			& when $i<O(\loglogε)$; \\
			O(2^{-i/4})	& otherwise.
		\end{cases*}\]*
		
		Now we recall a useful restatement of Fubini theorem
		in probability theory which states
		$𝔼[τ]=∑_{i=0}^∞ℙ\{τ>i\}$ \cite[Lemma~2.2.8]{Durrett10}.
		This reassure what we claimed above,
		that when does something first happen (LHS) is related to
		whether something happened before $i$ (RHS).
		The summation is from $i=0$ to $∞$ but we divide them into two cases:
		1)	form $i=0$ to $O(\loglogε)$, we have little control.
			We are summing $O(\loglogε)$ many $1$'s; the sum is $O(\loglogε)$.
		2)	for $i=O(\loglogε)$ to $∞$ we have the upper bound $O(2^{-i/4})$.
			We are summing a geometric series; the sum is $O(1)$.
		Put 1) and 2) together we have a complete estimate
		\begin{align*}
			𝔼[τ]
			&= ∑_{i=0}^∞ℙ\{τ>i\}=∑_{\text{1)}}ℙ\{τ>i\}+∑_{\text{2)}}ℙ\{τ>i\} \\
			&≤ ∑_{\text{1)}}1+∑_{\text{2)}}O(2^{-i/4})= O(\loglogε)+O(1) \\
			&= O(\loglogε).
		\end{align*}
		This closes the computation of $𝔼[τ]$.
	\end{IEEEproof}
	
	Theorem~\ref{thm:grow} contains the most technical steps in this work.
	This is the first time the concept of stopping time is introduced
	to the field of polar codes, and it plays key roles in the proof.
	Now we complete Theorem~\ref{thm:grow}, i.e.,
	the construction of the tree $𝒯$, it remains to:
	1)	read off the $N,\bC$ from $𝒯$;
	2)	define $𝒜$; and
	3)	read off the $P,R$ from $(𝒯,𝒜)$.
	
	\begin{prop} \label{prop:N}
		The tree $𝒯$ defined by Framed Rule~(\ref{eq:rule})
		possesses block length $N=2^n=ε^{-5}$.
	\end{prop}
	\begin{IEEEproof}
		Framed Rule~(\ref{eq:rule}) stops us from applying $\T$ at depth $n$.
		Thus it grows a tree of depth (at most) $n$,
		where $n$ was defined to be $-5\log_2ε$ in Theorem~\ref{thm:grow}.
		This leads to a code with block length $N$ (at most) $2^n=ε^{-5}$.
	\end{IEEEproof}
	
	\begin{prop} \label{prop:C}
		The tree $𝒯$ defined by Framed Rule~(\ref{eq:rule})
		possesses per-bit time complexity $\bC=O(\loglogε)$.
	\end{prop}
	\begin{IEEEproof}
		$𝔼[τ]$ is $O(\loglogε)$ by Theorem~\ref{thm:grow}.
		By the discussion that leads to Formula~(\ref{eq:Etau}),
		the per-bit time complexity $\bC$ is thus $𝔼[τ]=O(\loglogε)$.
	\end{IEEEproof}
	
	\begin{prop} \label{prop:P}
		Given $𝒯$ defined by Framed Rule~(\ref{eq:rule}),
		declare $𝒜$ by Criterion~(\ref{eq:util}), i.e.,
		\[*\TArule{
			$w∈𝒜$ if and only if \\
			$w$ is a leaf and $Z(w)≤ε2^{-n}$.
		}\]*
		Then $(𝒯,𝒜)$ possesses block error probability $ε$.
	\end{prop}
	\begin{IEEEproof}
		We compute the error probability as follows:
		\begin{align}
			P
			&≤ \sum_{w\in𝒜}Nℙ\{K_τ=w\}Z(w) \tag{union bound}\\
			&≤ \sum_{w\in𝒜}Nℙ\{K_τ=w\}ε2^{-n} \tag{Criterion~(\ref{eq:util})}\\
			&≤ Nε2^{-n}=ε. \tag{see below}
		\end{align}
		Here (see below) uses that $\{K_τ=w\}$ are disjoint events
		so their probability measures sum to $1$, at most.
		This proves the claim that $P≤ε$.
	\end{IEEEproof}
	
	\begin{prop} \label{prop:R}
		The pair $(𝒯,𝒜)$ defined by Framed Rule~(\ref{eq:rule})
		and Criterion~(\ref{eq:util}) possesses code rate $I(W)-ε$.
	\end{prop}
	\begin{IEEEproof}
		The sample space is partitioned into the following three events:
		\begin{align*}
			G &≔ \{0≤Z_τ≤ε2^{-n}\}; \\
			M &≔ \{ε2^{-n}<Z_i<1-ε2^{-n}\text{ for all }i≤n\}; \\
			B &≔ \{1-ε2^{-n}≤Z_τ≤1\}.
		\end{align*}
		Compare this to what we said after Criterion~(\ref{eq:util}).
		Event $G$ means $K_τ$ is a good channel; corresponding to 2).
		Event $M$ means $τ=n$ and $K_n$ is mediocre; corresponding to 4).
		Event $B$ means $K_τ$ is a bad channel; corresponding to 3).
		The second event $M$ is contained in $\{τ>n-1\}$
		(that $K_i$ is sufficiently polarized does not happen).
		By the proof of Theorem~\ref{thm:grow} we have
		\[*ℙ\{τ>n-1\}≤\begin{cases*}
			1				& if $n-1<O(\loglogε)$; \\
			O(2^{-(n-1)/4})	& otherwise.
		\end{cases*}\]*
		Recall $n≔-5\log_2ε$, so $n-1<O(\loglogε)$ does not happen as $ε→0$.
		The ``otherwise'' bound $O(2^{-(n-1)/4})=O(2^{-n/4})$ applies:
		\[ℙ(M)≤ℙ\{τ>n-1\}=O(2^{-n/4}). \label{eq:PMbound}\]
		
		Use this to rewrite the capacity as follows;
		here $𝕀(\bullet)$ is the indicator function of events:
		\begin{align}
			I(W)
			&= I_0=𝔼[I_τ] \tag{by (\ref{eq:conserv})}\\
			&= 𝔼[I_τ𝕀(G)]+𝔼[I_τ𝕀(M)]+𝔼[I_τ𝕀(B)] \tag{partition}\\
			&= 𝔼[I_τ𝕀(G)]+𝔼[I_τ𝕀(M)]+𝔼[(1-Z_τ)𝕀(B)] \tag{BEC}\\
			&≤ 𝔼[𝕀(G)]+𝔼[𝕀(M)]+ε2^{-n}𝔼[𝕀(B)] \tag{see below}\\
			&= ℙ(G)+ℙ(M)+ε2^{-n}ℙ(B) \tag{$𝔼𝕀$ is $ℙ$}\\
			&≤ ℙ(G)+O(2^{-n/4})+ε2^{-n}. \tag{by (\ref{eq:PMbound})}
		\end{align}
		Here (see below) is by $I_τ≤1$ for $G$ and $M$,
		and by $1-ε2^{-n}≤Z_τ$ for $B$.
		Use the last line to bound the code rate:
		\begin{align*}
			R
			&= ℙ\{K_τ∈𝒜\}=ℙ(G) \tag{Criterion~(\ref{eq:util})}\\
			&≥ I(W)-O(2^{-n/4})-ε2^{-n} \tag{rewrite $I(W)$}\\
			&= I(W)-O(ε^{5/4})-ε^6 \tag{$n≔-5\log_2ε$}\\
			&≥ I(W)-ε \tag{as $ε→0$}
		\end{align*}
		This proves the claim that $R≥I(W)-ε$.
	\end{IEEEproof}

\section{Connection to Other Works} \label{sec:connect}

\subsection{Pruned Codes in Terms of Deleting Vertexes}

	\cite{AYK11} introduces the so-called
	\emph{simplified successive cancellation} decoder, working as follows:
	During the construction of polar codes,
	some synthetic channel, for instance $(W♭)♭$, may find that
	all its descendants are frozen (potentially because $(W♭)♭$ is too bad).
	In such case, it is unnecessary to establish
	the part of the circuit that corresponds to its children.
	This results in circuits and trees like Fig.~\ref{fig:prune7}.
	
	\cite{AYK11} calls the synthetic channel $(W♭)♭$ a \emph{rate-zero node}.
	Similarly, a \emph{rate-one node} is a synthetic channel that is so good,
	all of its descendants being utilized.
	In such case, \cite{AYK11} argues that it could save some time
	by shortcutting the classical successive cancellation decoder.
	In particular, they turn soft-decision
	(calculation of \latin{a posteriori} probabilities)
	into hard-decision (XORing of bits).
	
	That said, we can save more by not applying $\T$ in the first place,
	ultimately reducing the per-bit time complexity~$\bC$
	from $O(\log N)$ to $O(\log\log N)$.
	We admit that this is not a fair comparison since \cite{AYK11} is aiming for
	practical performance while our result deals with mathematical asymptote.
	
	\cite{ZZWZP15} applies similar reduction to polar codes with other kernels.
	\cite{ZZPYG14} gives a similar approach, but is based on belief propagation.

\subsection{Pruned Codes in Terms of Adding Vertexes}

	\cite{EKMFLK15,EKMFLK17} introduce the so called ``relaxed polarization.''
	\cite{WLZZ15} introduces the so-called ``selective polarization.''
	They suggest that when some synthetic channel, say $(W♭)♯$,
	is not perfectly polarized,
	it should be further polarized by concatenating with an outer polar code.
	This results in trees like Fig.~\ref{fig:prune5}.
	It is worth noting that \cite{EKMFLK17}
	attempts to compute the saving in time mathematically.
	Since they want the final $P$ be $2^{-2^{βn}}$ for some $β<1/2$,
	every channel must undergo at least $βn$ rounds of $\T$
	to square its \Bha{} parameter that many times.
	Thus their final $\bC$ is still $Ω(βn)=Ω(\log N)$, not any lower.
	
	\cite{EECB17} illustrate another attempt,
	which they called ``code augmentation,''
	to protect unpolarized channels by appending polar codes to them.
	\cite{WYY18,WYXY18} do very similar things
	which they called ``information-coupling.''
	They protect unpolarized channels by
	repeating the same symbol across several code blocks 

\subsection{Relation to Special Treatment of Subtrees}

	Recall the recursive definition
	\[*Z_{i+1}=\begin{cases*}
		1-(1-Z_i)^2 & w.p.\ $1/2$ (head); \\
		Z_i^2       & w.p.\ $1/2$ (tail).
	\end{cases*}\]*
	Assume there is some $Z_m$ such that
	$ε2^{-n}<Z_m<ε2^{m-7n/5}$ for some $m∈[2n/5,n]$.
	It is clear that although this synthetic channel is quite good,
	it is not good enough to become a leaf.
	(At least in terms of Criterion~\ref{eq:util}.)
	What can we say about its descendants?
	
	Since $Z_{m+i+1}<2Z_{m+i}$,
	it turns out $Z_{m+i}<2^iZ_m<ε2^{i+m-7n/5}<ε2^{-2n/5}$ for all $i<n-m$.
	Thus if tail ever happens, say at time $m+i+1$,
	then $Z_{m+i+1}=Z_{m+i}^2<ε^22^{-4n/5}=ε2^{-n}$, which means a leaf.
	That is, the subtree rooted at $K_m$ is such that
	every lower child becomes a leaf, and
	every upper child has children, till depth $n$.
	The upper child at depth $n$ is then frozen
	while all other leaves are utilized.
	See Fig.~\ref{fig:fast} for visualization.
	
	\begin{figure}
		$$\tikz[ut]{
			\utnode$K_m$\T${
				\utnode$K_m♭$\T${
					\utnode$(K_m♭)♭$\T${
						\utnode$((K_m♭)♭)♭\mathrlap\dotso$${}{} }{
						\utnode$((K_m♭)♭)♯\mathrlap\dotso$${}{} } }{
					\utnode$(K_m♭)♯$${}{} } }{
				\utnode$K_m♯$${}{} }
		}\qquad\tikz[ut]{
			\utnode$K_m$\T${
				\utnode$K_m♭$${}{} }{
				\utnode$K_m♯$\T${
					\utnode$(K_m♯)♭$${}{} }{
					\utnode$(K_m♯)♯$\T${
						\utnode$((K_m♯)♯)♭\mathrlap\dotso$${}{} }{
						\utnode$((K_m♯)♯)♯\mathrlap\dotso$${}{} } } }
		}\quad$$
		\caption{
			On the left is the subtree rooted at $K_m$
			if $ε2^{-n}<Z_m<ε2^{m-7n/5}$ for some $m∈[2n/5,n]$.
			Every time coin toss selects the lower child
			the child has its \Bha{} parameter squared.
			The child's \Bha{} parameter is small enough so $\T$ is not applied.
			This makes it a leaf.
			At the end, the leaf $(\dotso(K_m♭)♭\dotso)♭$ is frozen.
			One the right is the subtree rooted at $K_m$
			if $ε2^{-n}<1-Z_m<ε2^{m-7n/5}$ for some $m∈[2n/5,n]$.
			Every time coin toss selects the upper child
			the child has its capacity squared.
			The child's \Bha{} parameter is large enough so $\T$ is not applied.
			This makes it a leaf.
		}
		\label{fig:fast}
	\end{figure}
	
	\cite{SG13} recognizes that this subtree
	generates a single-parity-check subcode,
	which can be decoded more efficiently than the butterfly devices do.
	
	Similarly, a $Z_m$ that is close enough to the top threshold $1-ε2^{-n}$
	generates a subtree that mainly ``grows downward''
	and every leaf except the very bottom one is frozen.
	This either induces a trivial code (if the very bottom leaf is frozen)
	or a repetition code (if the very bottom leaf is utilized)
	Again, repetition codes can be efficiently decoded.
	See Fig.~\ref{fig:fast} for visualization.
	
	The simulation by \cite{SG13}, and subsequently by \cite{SGVTG14},
	suggests that this \latin{ad hoc} treatment
	accelerates the real world performance.
	For our purpose, however,
	special treatment makes it difficult to describe the complexity.

\subsection{Motivation from Systematic Polar Coding}

	\cite{Arikan11} suggests systematic polar coding,
	where the receiver is not interested in $ˆu$
	but wants to recover $x$ from $y$.
	One consequence is that, if the two right pins of the butterfly device
	\tikz[inline]\pic{D}; correspond to two frozen channels,
	then this device can be dropped without affecting
	the overall decoding ability of the circuit.
	Similarly, if the two right pins correspond to two utilized channels,
	it could also be dropped.
	
	The argument above gives another reason (or perspective)
	why the tree should be pruned.
	One may keep dropping butterfly devices (keep pruning the tree)
	till it stabilizes.
	It is easy to see that a device remains if and only if
	some of its children are frozen and some are utilized.
	Our intuition suggests that the number of remaining devices is
	\[O（N\log\Bigl\lvert\log\frac{ϵ}N\Bigr\rvert）\]
	where $ϵ$ is the threshold of a channel being utilized
	(which is $ε2^{-n}$ in our construction).
	When $N$ is polynomial in $ϵ$, this reassures out result.
	
	Therefore, that $\bC$ is log-logarithm in $N,R,P$ also follows if one
	applies systematic polar coding \cite{Arikan11}
	with simplified successive cancellation decoding \cite{AYK11},
	and then analyzes the performance using \cite{GX13} or \cite{MHU16}.

\section{Symmetric \texorpdfstring{$q$}{q}-ary Memoryless Channels}
	\label{sec:qchannel}

	In this section we generalize Theorem~\ref{thm:main}.
	Fix a prime $q$.
	Fix a $q$-ary-input discrete symmetric memoryless channel $W$.
	We will show that an analog of Theorem~\ref{thm:main} holds for $W$.
	
	In this setting, \Arikan's $\llol$ kernel ``still works.''
	By still working we mean the definitions of  circuit, tree $𝒯$,
	transformation $\T$, and processes $K_i,I_i,K_{i∧τ},I_{i∧τ}$ still apply.
	That $I_i,I_{i∧τ}$ are martingales still holds.
	The phenomenon that channels polarize is preserved, i.e.,
	$\lim_{i→∞}I_i∈\{0,1\}$ when $I$ is normalized \cite[Corollary~15]{MT14}.
	Not only do notations make sense,
	but also the proof we presented above is (almost) sound.
	That is, we can almost claim that
	$(N,R,P,\bC)$ is $\(ε^{-5},I(W)-ε,ε,O(\loglogε)\)$ except that
	Estimate~(\ref{eq:Y24O4}) does not hold in the first place.
	To that end, we need the following substitute of Estimate~(\ref{eq:Y24O4}).
	
	\begin{thm} \label{thm:qexp}
		For any prime $q$ and
		any $q$-ary-input discrete symmetric memoryless channel,
		there exist constants $μ>0$ and $β>0$ such that
		the process $Y_i≔\min\{I_i,1-I_i\}$ satisfies
		\[*ℙ｛Y_i>2^{-2^{βi}}｝≤O(2^{-i/μ}).\]*
	\end{thm}
	\begin{IEEEproof}
		The proof is deferred until Appendix~\ref{pf:qexp}.
		But it is worth mentioning that \cite{BGNRS18,BGS18} inspire us.
		In particular, it is \cite[Lemma~6.3]{BGNRS18}
		that makes up the last piece of the puzzle.
	\end{IEEEproof}
	
	Once we have the substitution of Estimate~(\ref{eq:Y24O4})
	the general strategy is to repeat Theorem~\ref{thm:grow}
	and then repeat Propositions \ref{prop:N} to~\ref{prop:R}.
	But we need the following modification:
	\begin{itemize}
		\item $Y(w)$ becomes $\min\{I(w),1-I(w)\}$.
		\item $n≔-5\log_2ε$ is replaced by $n≔-2μ\log_2ε$.
		\item $4$ is replaced by $μ$.
		\item $1/24$ is replaced by $β$.
		\item It is no longer true that $I(W)+Z(W)=1$.
	\end{itemize}
	Although $I(W)+Z(W)=1$ is not true anymore,
	$1-I(W)$ and $Z(W)$ are ``bi-Hölder''%
	\footnote{
		``Bi-Hölder'' is a temporary name
		inspired by the bi-Lipschitz condition and the Hölder condition.
		It is denoted as ``$A\stackrel e\sim B$'' in \cite{MT14}.
	}
	in the sense that $aZ(W)^b≤1-I(W)≤cZ(W)^d$
	for some positive constants $a,b,c,d$ depending on $q$ but not $W$.
	In a looser language, $I(W)$ and $Z(W)$ control each other polynomially
	when $\(I(W),Z(W)\)$ is close to $(0,1)$.
	Since we expect $I_i$ or $Z_i$ to converge to $1$ or $0$
	(doubly) exponentially fast, the polynomial factor does not matter.
	They \emph{both} converge to $0$ or $1$ (doubly) exponentially fast.
	Besides, \cite{MT14} defines three extra measurements
	$P_e(W)$, $T(W)$, and $S(W)$.
	All five $I,Z,P_e,T,S$ are mutually ``bi-Hölder'' up to rescaling.
	See \cite[Definition~27 and Corollary~28]{MT14} for details.%
	\footnote{
		\cite[Corollary~28]{MT14} covers $Z,P_e,T,Z$ but not $I$.
		But $I$ and $P_e$ are related to each other
		by Fano's inequality and its converse.
	}
	
	Here is the precise statement of
	the generalization of Theorem~\ref{thm:main}.
	
	\begin{thm} \label{thm:qmain}
		For any prime $q$ and
		any $q$-ary-input discrete symmetric memoryless channel,
		there exists a constant $μ$ such that, for small $ε>0$,
		there are codes with
		block length~$ε^{-3μ}$, code rate~$I(W)-ε$, error probability~$ε$, and
		encoding and decoding time complexity $O(\loglogε)$ per information bit.
	\end{thm}
	\begin{IEEEproof}
		Analog of Theorem~\ref{thm:grow}:
		We will show that Framed Rule~(\ref{eq:rule}), i.e.,
		\[*\TArule{
			Apply $\T$ to $w$ if and only if \\
			$\depth(w)<n$ and $Y(w)>ε2^{-n}$,
		}\]*
		generates a channel tree with $𝔼[τ]=O(\loglogε)$.
		
		Note that $Y(w)$ in the rule became $\min\{I(w),1-I(w)\}$,
		and the new $n$ is $-2μ\log_2ε$.
		By the same reason presented in the proof of Theorem~\ref{thm:grow},
		\[*\{τ>i\}⊂\{Y_i>ε2^{-n}\}=\{Y_i>ε^{1+2μ}\}.\]*
		This again can be divided into
		$Y_i>2^{-2^{βi}}$ or $2^{-2^{βi}}>ε^{1+2μ}$.
		The first disjunct is controlled by Theorem~\ref{thm:qexp}.
		The second disjunct becomes $i<O(\loglogε)$.
		So we have a joint bound
		\[*ℙ\{τ>i\}≤\begin{cases*}
			1			& when $i<O(\loglogε)$; \\
			O(2^{-i/μ})	& otherwise.
		\end{cases*}\]*
		By Fubini \cite[Lemma~2.2.8]{Durrett10},
		\[𝔼[τ]=∑_{i=0}^∞ℙ\{τ>i\}=O(\loglogε). \label{eq:qlogloge}\]
		This finishes the computation of $𝔼[τ]$.
		
		Analog of Proposition~\ref{prop:N}:
		The Framed Rule~\ref{eq:rule} stops us from applying $\T$ at depth $n$.
		Thus the block length is $N=2^n=ε^{-2μ}$.
		Remark: this is not a typo, we do want $ε^{-2μ}$ instead of $ε^{-3μ}$.
		
		Analog of Proposition~\ref{prop:C}:
		By Estimate~(\ref{eq:qlogloge}),
		the tree $𝒯$ defined by Framed Rule~(\ref{eq:rule})
		possesses per-bit time complexity $𝔼[τ]=O(\loglogε)$.
		
		Analog of Proposition~\ref{prop:P}:
		Given $𝒯$ defined by Framed Rule~(\ref{eq:rule}), declare $𝒜$ by
		\[\TArule{
			$w∈𝒜$ if and only if \\
			$w$ is a leaf and $1-I(w)≤ε2^{-n}$.
		} \label{eq:qutil}\]
		Then $(𝒯,𝒜)$ possesses block error probability $εO(1)$.
		Here is the calculation:
		\begin{align}
			P
			&≤ \sum_{w\in𝒜}Nℙ\{K_τ=w\}(1-I(w))O(1) \tag{see below}\\
			&≤ \sum_{w\in𝒜}Nℙ\{K_τ=w\}ε2^{-n}O(1)
											\tag{Criterion~(\ref{eq:qutil})}\\
			&≤ Nε2^{-n}O(1)≤εO(1). \tag{sigma-additivity}
		\end{align}
		Here (see below) is a two-step bound:
		First by union bound, the total error probability is at most
		the sum of error probabilities of individual channels in $𝒜$.
		To bound the later, we recall \cite[Theorem~1 and Formula~(14)]{FM94}.
		The theorem therein implies that
		the error probability of $w$ is linear in $1-I(w)$.
		Hence the bound $(1-I(w))O(1)$.
		The resulting bound $P≤εO(1)$ is good enough
		and we will live with it, temporarily.
		
		Analog of Proposition~\ref{prop:R}:
		we will show that $(𝒯,𝒜)$ defined above possesses code rate $I(W)-ε$.
		
		The sample space is partitioned into the following three events:
		\begin{align*}
			G &≔ \{1-ε2^{-n}≤I_τ≤1\}; \\
			M &≔ \{ε2^{-n}<I_i<1-ε2^{-n}\text{ for all }i≤n\}; \\
			B &≔ \{0≤I_τ≤ε2^{-n}\}.
		\end{align*}
		The second event is contained in $\{τ>n-1\}$, where
		\[*ℙ\{τ>n-1\}≤\begin{cases*}
			1			& if $n-1<O(\loglogε)$; \\
			O(2^{-(n-1)/μ})	& otherwise.
		\end{cases*}\]*
		As $ε→0$ and $n→∞$ we do not expect $n-1<O(\loglogε)$, so
		\[ℙ(M)≤ℙ\{τ>n-1\}=O(2^{-n/μ}). \label{eq:qPMbound}\]
		Use this to rewrite the capacity as follows
		\begin{align}
			I(W)
			&= I_0=𝔼[I_τ] \tag{by generalized (\ref{eq:conserv})}\\
			&= 𝔼[I_τ𝕀(G)]+𝔼[I_τ𝕀(M)]+𝔼[I_τ𝕀(B)] \tag{partition}\\
			&≤ 𝔼[𝕀(G)]+𝔼[𝕀(M)]+ε2^{-n}𝔼[𝕀(B)] \tag{see below}\\
			&= ℙ(G)+ℙ(M)+ε2^{-n}ℙ(B) \tag{$𝔼𝕀=ℙ$}\\
			&≤ ℙ(G)+O(2^{-n/μ})+ε2^{-n}. \tag{by (\ref{eq:qPMbound})}
		\end{align}
		Here (see below) is by $I_τ≤1$ for $G$ and $M$,
		and by $I_τ≤ε2^{-n}$ for $B$.
		Use the last line to bound the code rate:
		\begin{align*}
			R
			&= ℙ\{K_τ∈𝒜\}=ℙ(G) \tag{Criterion~(\ref{eq:qutil})}\\
			&≥ I(W)-O(2^{-n/μ})-ε2^{-n} \tag{rewrite $I(W)$}\\
			&= I(W)-O(ε^2)-ε^{1+2μ} \tag{$n≔-2μ\log_2ε$}\\
			&≥ I(W)-ε \tag{as $ε→0$}
		\end{align*}
		This proves the claim that $R≥I(W)-ε$.
		
		So far we proved that there are codes with
		$(N,R,P,\bC)=\(ε^{-2μ},I(W)-ε,εO(1),O(\loglogε)\)$.
		We do not like the extra $O(1)$ term.
		So we replace $ε$ by $ε/O(1)$ to obtain
		$\(Ω(ε^{-2μ}),I(W)-Ω(ε),ε,O(\loglogε)\)$.
		This can be loosened to $\(ε^{-3μ},I(W)-ε,ε,O(\loglogε)\)$ given $ε→0$.
		This finishes the proof.
	\end{IEEEproof}
	
	\begin{figure}
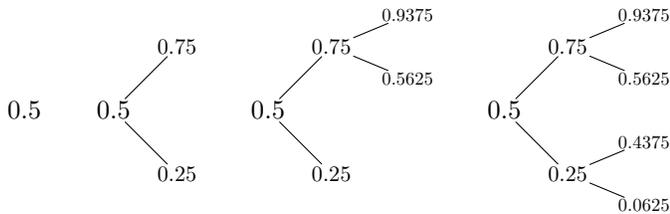

		\def\qhfill{\hskip \textwidth minus \textwidth }
		$$\tikz[ut]{
			\utnode$0.5$${}{}
		}\qhfill\tikz[ut]{
			\utnode$0.5$${
				\utnode$0.75$${}{} }{
				\utnode$0.25$${}{} }
		}\qhfill\tikz[ut]{
			\utnode$0.5$${
				\utnode$0.75$${
					\utnode$0.9375$${}{} }{
					\utnode$0.5625$${}{} } }{
				\utnode$0.25$${}{} }
		}\qhfill\tikz[ut]{
			\utnode$0.5$${
				\utnode$0.75$${
					\utnode$0.9375$${}{} }{
					\utnode$0.5625$${}{} } }{
				\utnode$0.25$${
					\utnode$0.4375$${}{} }{
					\utnode$0.0625$${}{} } }
		}$$
		\caption{
			Step one on the left:
			Start with $W$ and write down $Z(W)$, which is $0.5$.
			Step two the second from the left:
			Both $0.5$ and $1-0.5$ are larger than $ε2^{-n}$.
			Apply $\T$ to $0.5$ to obtain two synthetic channels
			$1-(1-0.5)^2=0.75$ and $0.5^2=0.25$.
			Append them as children of $0.5$.
			Step three the second from the right:
			Both $0.75$ and $1-0.75$ are larger than $ε2^{-n}$.
			Apply $\T$ to $0.75$ to obtain two synthetic channels
			$1-(1-0.75)^2=0.9375$ and $0.75^2=0.5625$.
			Append them as children of $0.75$.
			Step four on the right:
			Both $0.25$ and $1-0.25$ are larger than $ε2^{-n}$.
			Apply $\T$ to $0.25$ to obtain two synthetic channels
			$1-(1-0.25)^2=0.4375$ and $0.25^2=0.0625$.
			Append them as children of $0.25$.
		}
		\label{fig:grow1234}
	\end{figure}

\section{Future Works}

	We are not satisfied by our generalized result Theorem~\ref{thm:qmain}
	for two reasons:
	1)	Its constants $β,μ$ depend on the channel $W$.
		The dependency comes from Theorem~\ref{thm:qexp},
		but for general channels we know very little.
	2)	It applies to prime $q$ but not prime powers.
		We hope this can be generalized to at least prime powers.
		Once done, we can hope for all discrete-input channels.
	
	From studies of random codes,
	$I(W)-R$ is polynomial in $N$ while $P$ is exponential in $N$.
	Thus it seems improper to parametrize
	$I(W)-R$ and $P$ with a single variable $ε$.
	It would be interesting if one could come up with a description
	of more general trade-offs among $N$, $R$, $P$, and time complexity.

\section{Conclusions}

	We proposed a pruned variant of polar coding where
	the channel tree is pruned by closely looking at the \Bha{} parameters.
	We proved that the resulting per-bit complexity is log-logarithmic
	in block length, in gap to capacity and in error probability.
	This constitutes the only family of codes possessing this property.
	
	Similar ideas have appeared in existing works
	mentioned in Section~\ref{sec:connect}, namely
	simplified successive cancellation decoder, relaxed polarization,
	selective polarization, code augmentation, and information-coupling.
	They found that doing this type of simplification
	reduces the wall-clock time of coding significantly.
	Alongside their success, we prove for the first time
	the log-logarithmic asymptote for polar codes.
	
	In spite of the fact that the log-logarithmic asymptote
	is not record-breaking as other constructions with
	bounded per-bit complexity $\bC$ exist (\cite{PSU05,PS07}),
	the log-logarithmic asymptote is the second best thing after boundedness.
	Besides, our construction takes block length $N$ in to consideration.
	There are only two families of capacity-achieving codes whose $N$
	is explicitly characterized together with $R$, $P$, and $\bC$:
	random coding and classical polar coding.
	Our construction gives the third family.
	
	Finally we generalize our result to $q$-ary symmetric channels
	where the log-logarithmic asymptote of polar codes
	becomes the lowest per-bit complexity known to date.
	This suggests that the log-logarithmic asymptote is rather
	a universal behavior not limited to BEC,
	just like channel polarization is
	a universal phenomenon on all discrete channels.
	We look forward to generalization of
	our result to all  discrete symmetric channels.

\appendix

\subsection{Execution of Framed Rule (\ref{eq:rule})} \label{app:grow}

	\begin{figure}
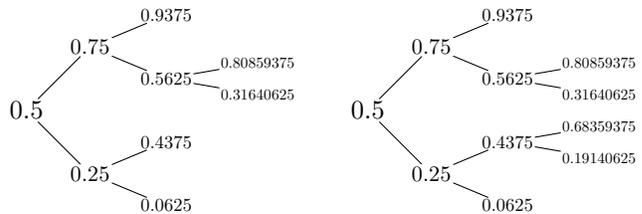

		$$\tikz[ut]{
			\utnode$0.5$${
				\utnode$0.75$${
					\utnode$0.9375$${}{} }{
					\utnode$0.5625$${
						\utnode$0.80859375$${}{} }{
						\utnode$0.31640625$${}{} } } }{
				\utnode$0.25$${
					\utnode$0.4375$${}{} }{
					\utnode$0.0625$${}{} } }
		}\kern2em\tikz[ut]{
			\utnode$0.5$${
				\utnode$0.75$${
					\utnode$0.9375$${}{} }{
					\utnode$0.5625$${
						\utnode$0.80859375$${}{} }{
						\utnode$0.31640625$${}{} } } }{
				\utnode$0.25$${
					\utnode$0.4375$${
						\utnode$0.68359375$${}{} }{
						\utnode$0.19140625$${}{} } }{
					\utnode$0.0625$${}{} } }
		}$$
		\caption{
			Step five and six on the left:
			$1-0.9375$ is smaller than $ε2^{-n}$.
			Do not apply $\T$; let $0.9375$ be a leaf.
			Both $0.5625$ and $1-0.5625$ are larger than $ε2^{-n}$.
			Apply $\T$ to $0.5625$ to obtain two synthetic channels
			$1-(1-0.5625)^2=0.80859375$ and $0.5625^2=0.31640625$.
			Append them as children of $0.5625$.
			Step seven and eight on the right:
			Both $0.4375$ and $1-0.4375$ are larger than $ε2^{-n}$.
			Apply $\T$ to $0.4375$ to obtain two synthetic channels
			$1-(1-0.4375)^2=0.68359375$ and $0.4375^2=0.19140625$.
			Finally $0.0625$ is smaller than $ε2^{-n}$.
			Do not apply $\T$; let $0.0625$ be a leaf.
			Now we reach depth $n=3$; terminate.
		}
		\label{fig:grow5678}
	\end{figure}
	
	We present a possible execution of Framed Rule~(\ref{eq:rule}), i.e.,
	\[*\TArule{
		Apply $\T$ to $w$ if and only if \\
		$\depth(w)<n$ and $Y(w)>ε2^{-n}$.
	}\]*
	Let $Z(W)=0.5$; let $ε=0.8$.
	We should have calculated $n$ by $ε$; but we choose $n=3$ for simplicity.
	Note that $ε2^{-n}=0.1$.
	Also we should have applied $\T$ to \emph{channels}.
	But for BEC, the \Bha{} parameter uniquely determines the channel;
	Thus by applying $\T$ to a number $a$ to obtain other numbers $b,c$,
	we meant to apply $\T$ to BEC of erasure probability $a$
	to obtain BECs of erasure probabilities $b,c$.
	See Fig.~\ref{fig:grow1234} for steps one to four.
	See Fig.~\ref{fig:grow5678} for steps five to eight.

\subsection{Numerical Evidence of Inequality~(\ref{eq:plug424})}
	\label{app:numerical}

\begin{figure}
		\pgfmathdeclarefunction{h2}{1}{%
			\pgfmathparse{-#1*log2(#1)-(1-#1)*log2(1-#1)}}
		\tikz{
			\begin{axis}[ylabel=LHS of Inequality~(\ref{eq:plug424}),
				         xlabel={$π$}]
				\addplot[domain=0:.9,samples=40]
					{(1-\x)/(4-3.627*\x)+h2(1/(24-21.762*\x))};
				\addplot[domain=.9:1,samples=100]
					{(1-\x)/(4-3.627*\x)+h2(1/(24-21.762*\x))};
			\end{axis}
		}
		\caption{
			This is an attempt to verify Inequality~(\ref{eq:plug424})
			by plotting the LHS of the inequality.
			The plot is done in \TeX{} so we expect rounding errors.
			The plot shows that for $π∈[0.9,1]$
			the LHS is very close to $1$.
			This suggests that the choice of 
			constants $μ',β'$ is close the optimal.
			To verify the inequality more rigorously,
			we divide the interval $[0.9,1]$ into $100000$ subintervals
			and use interval arithmetic to \emph{prove} inequality.
			Remark: Later computation shows that the local maximum is at
			$(0.999930450125367, 0.9864109898636828)$.
		}
		\label{fig:numerical}
	\end{figure}
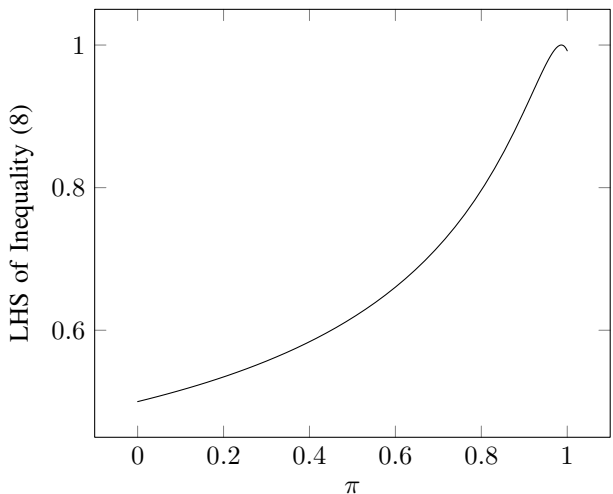

	We want to verify Inequality~(\ref{eq:plug424}), i.e.,
	\[*\frac4{4-3.627π}+h_2（\frac{1/6}{4-3.627π}）<1,\]*
	for all $\pi∈[0,1]$.
	See Fig.~\ref{fig:numerical} for an approximated plot.
	We see that when $0.9<π<1$ it is difficult to tell
	whether LHS of (\ref{eq:plug424}) is smaller than $1$ or not.
	We decide to verify this using interval arithmetic.
	Interval arithmetic treats an interval as a number with uncertainty.
	When rounding takes place, it rounds toward the \emph{safe} direction.
	For instance, $\cos([3.14,3.15])$ returns
	$[-1.0000000000000000,-0.99996465847134186]$
	instead of $[\cos(3.14),\cos(3.15)]$.
	We write a SageMath script that:
	1) divides the interval $[0.9,1]$ into $100000$ subintervals evenly; and
	2) for every subinterval checks if the inequality holds.
	All subintervals pass the check.
	Remark:
		dividing $[0.9,1]$ into $10000$ subintervals
		does not verify the inequality.
		This is because the arithmetic rounds upward so much
		that eventually the upper bound becomes greater than $1$,
		which is not contradicting but inconclusive.
	
\subsection{Proof of Theorem~\ref{thm:qexp}} \label{pf:qexp}

	Fix a prime $q$.
	Fix a $q$-ary-input discrete symmetric memoryless channel $W$.
	We want to find constants $μ>0$ and $β>0$ such that
	the process $I_i$ satisfies
	\begin{gather*}
		ℙ｛I_i≤2^{-2^{βi}}｝≥1-I(W)-O(2^{-i/μ}), \\
		ℙ｛1-I_i≤2^{-2^{βi}}｝≥I(W)-O(2^{-i/μ}).
	\end{gather*}
	We borrow terminologies and lemmas from \cite{BGNRS18}
	for a head start.
	
	By \cite[Definition~1.8]{BGNRS18},
	the matrix $\llol$ is mixing.
	By \cite[Theorem~1.10]{BGNRS18},
	the process $I_i$ corresponding to $\llol$ is locally polarizing.
	By \cite[Theorem~1.6]{BGNRS18},
	the process $I_i$ corresponding to $\llol$ is strongly polarizing.
	By \cite[Definition~1.4]{BGNRS18},
	the process $I_i$ is such that for all $γ>0$ there exist $η<1$ and $β'<∞$
	such that $I_i$ is $(γ^i,β'η^i)$-polarizing.
	By \cite[Definition~1.2]{BGNRS18}, $I_i$ is such that for all $γ>0$
	there exist $η<1$ and $β'<∞$ such that $ℙ\{I_i∈(γ^i,1-γ^i)\}<β'η^i.$
	
	Choose $γ=1/2$.
	We obtain:
	there exists $η<1$ such that $ℙ\{I_i∈(2^{-i},1-2^{-i})\}<O(η^i)$.
	Since $η<1$, the right hand side $O(η^i)$
	converges to $0$ exponentially fast.
	This means that the majority of $I_i$ are either
	exponentially small (i.e., $0≤I_i≤2^{-i}$) or
	exponentially close to $1$ (i.e., $1-2^{-i}≤I_i≤1$).
	What we want to show consists of two parts:
	1)	The proportion of $I_i$ that is
		exponentially small is about $1-I(W)$;
		the proportion of $I_i$ that is
		exponentially close to $1$ is about $I(W)$.
	2)	Exponentially small $I_i$'s are basically doubly-exponentially small
		(i.e., $0≤I_i≤2^{-2^{βi}}$);
		the close-to-$1$ counterpart is doubly-exponentially close to $1$
		(i.e., $1-2^{-2^{βi}}≤I_i≤1$).
	
	Now we go for 1).
	Observation: the result we want to prove and the tool we have in hand
	are symmetric in $I_i$ and $1-I_i$.
	It suffices to show, say, the close-to-$1$ part of the statement.
	The small-$I_i$ part follows by symmetry.
	
	Now we show $ℙ\{1-2^{-i}≤I_i≤1\}≥I(W)-O(η^i)$.
	Similar to Proposition~\ref{prop:R},
	we partition the sample space into three events
	\begin{align*}
		G &≔ \{1-2^{-i}≤I_i≤1\}; \\
		M &≔ \{2^{-i}<I_i<1-2^{-i}\}; \\
		B &≔ \{0≤I_i≤2^{-i}\}.
	\end{align*}
	Then $ℙ(M)=ℙ\{I_i∈(2^{-i},1-2^{-i})\}=O(η^i)$.
	Next we rewrite the capacity
	\begin{align}
		I(W)
		&= I_0=𝔼[I_i] \tag{martingale}\\
		&= 𝔼[I_i𝕀(G)]+𝔼[I_i𝕀(M)]+𝔼[I_i𝕀(B)] \tag{partition}\\
		&≤ 𝔼[𝕀(G)]+𝔼[𝕀(M)]+2^{-i}𝔼[𝕀(B)] \tag{see below}\\
		&= ℙ(G)+ℙ(M)+2^{-i}ℙ(B) \tag{$𝔼𝕀=ℙ$}\\
		&≤ ℙ(G)+O(η^i)+2^{-i}. \notag
	\end{align}
	Here (see below) is by $I_i≤1$ for $G$ and $M$,
	and by $I_i≤2^{-i}$ for $B$.
	Already we have that $ℙ\{1-2^{-i}≤I_i≤1\}=ℙ(G)≥I(W)-O(η^i)-O(2^{-i})$.
	We may assume $η>1/2$.
	Thus $ℙ\{1-2^{-i}≤I_i≤1\}≥I(W)-O(η^i)$.
	The flipped version $ℙ\{0≤I_i≤2^{-i}\}≥1-I(W)-O(η^i)$
	also holds by symmetry.
	This finishes the 1) part.
	
	Now we go for the small-$I_i$ part of 2).
	We need a lemma.
	By \cite[Lemma~6.3]{BGNRS18}, there exists a constant $Q>0$ such that
	$I_{i+1}≤QI_i^2$ if $K_{i+1}$ is the lower child.
	Clearly $I_i<1/Q^2$ implies $QI_i^2≤I_i^{1.5}$.
	So we deduce that
	whenever $I_i<1/Q^2$ and $K_{i+1}$ is the lower child, $I_{i+1}≤I_i^{1.5}$.
	Another case is when $K_{i+1}$ is the upper child.
	We choose a larger $Q$ such that $Q≥2^5$.
	Then whenever $I_i<1/Q^2$ and $K_{i+1}$ is the upper child,
	$I_{i+1}≤2I_i≤I_i^{0.9}$.
	Combine the two cases of $I_{i+1}$,
	we find that if $I_i<1/Q^2$ then $I_{i+1}$ is (at most)
	$I_i^{1.5}$ or $I_i^{0.9}$, each with probability $1/2$.
	We conclude this paragraph by rewriting this formally: when $I_i<1/Q^2$,
	\[*
		I_{i+1}≤
		\begin{cases*}
			I_i^{0.9} & w.p.\ $1/2$ (upper child case) \\
			I_i^{1.5} & w.p.\ $1/2$ (lower child case) .
		\end{cases*}
	\]*
	
	Now let $n$ be a large number.
	We know $ℙ\{0≤I_n≤2^{-n}\}≥1-I(W)-O(η^n)$.
	Now we continue the process for $i=n,\dotsc,4n$.
	We want to show that at step $4n$, the bad channels
	have doubly-exponentially small capacity.
	That is, we want $ℙ｛0≤I_{4n}≤2^{-2^{4βn}}｝≥1-I(W)-O(η^{4n})$ for some $β$.
	There are two obstacles:
	a)	If $I_i≥1/Q^2$, we lose control on $I_{i+1}$.
		We want to avoid this.
	b)	Even if $I_i<1/Q^2$, we want $I_i$ to go through
		the $1.5$-th power instead of the $0.9$-th power.
	We let $A$ be the event that $I_n<2^{-n}$ but $I_i>1/Q^2$ for some $n<i<4n$.
	When that happens, let $σ$ be the lowest $i$ such that $I_i>1/Q^2$.
	When that does not happen, we let $σ$ be $4n$.
	We let $B$ be the event that among $3n$ chances,
	$I_i$ undergoes the $1.5$-th power less than $n$ times.
	We now control $A$ and $B$.
	
	For $A$, we have $ℙ(A)=ℙ\{I_σ≥1/Q^2\}≤𝔼[I_σ]Q^2≤𝔼[I_n]Q^2≤Q^22^{-n}$
	by \cite[Theorem~5.7.6]{Durrett10}.
	For $B$, by Hoeffding's inequality,
	we may enlarge $η<1$ such that $ℙ(B)<O(η^n)$.
	We see that both $A$ and $B$ are rare events.
	In fact, their probability measures are both in $O(η^n)$.
	
	Finally we look at what happens outside $A∪B$:
	If $I_n<2^{-n}$ and neither $A$ or $B$ happens,
	then $I_i$ undergoes the $1.5$-th power $n$ times, at least;
	and undergoes the $0.9$-th power $2n$ times, at most.
	Thus $I_{4n}$ is at most $I_n$ to the $(1.5^n·0.9^{2n})$-th power.
	The exponent $1.5^n·0.9^{2n}$ is at least $2^{0.28n}$,
	so $I_{4n}≤(2^{-n})^{2^{0.28n}}≤2^{-2^{0.28n}}=2^{-2^{0.07·4n}}$.
	
	We review what we have so far:
	First the probability that $0≤I_n≤2^{-n}$ is at least $1-I(W)-O(η^n)$.
	And then we continue the process for $i=n,\dotsc,4n$.
	We lose some $I_i$ in $A$; this costs us $Q^22^{-n}$.
	We lose some $I_i$ in $B$; this costs us $O(η^n)$.
	As $n→∞$ the constant $Q$ does not matter; we lose $2O(η^n)$.
	What are left are some $I_i$ such that $I_{4n}≤2^{-2^{0.07·4n}}$.
	Therefore, we have just proven that
	$ℙ｛I_{4n}≤2^{-2^{0.07·4n}}｝≤1-I(W)-3O(η^n)$.
	Now we choose $μ>0$ and $β>0$ such that
	$ℙ｛I_{4n}≤2^{-2^{4βn}}｝≥1-I(W)-O(2^{-4n/μ})$.
	This finishes the small-$I_i$ part of 2).
	
	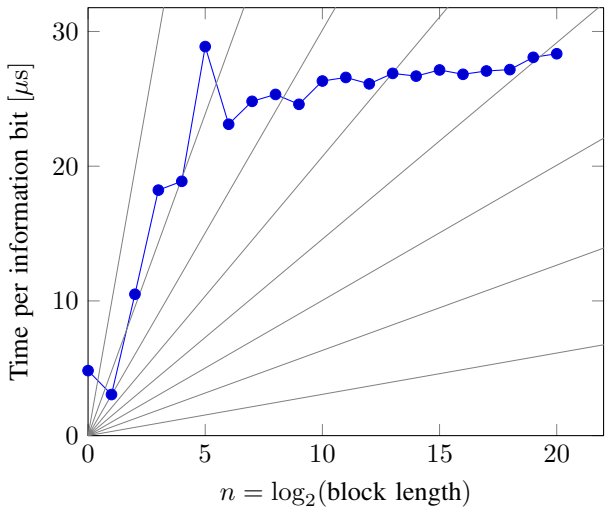
\begin{figure}
		\tikz{
			\begin{axis}[simu,
				         ylabel=Time per information bit,
				         change y base,y SI prefix=micro,y unit=s,
				         xlabel={$n=\log_2(\text{block length})$}]
				\addplot table{
					0	4.82528412249e-06
					1	3.04982745547e-06
					2	1.04944288069e-05
					3	1.8223561104e-05
					4	1.88777563987e-05
					5	2.88788760372e-05
					6	2.31143033792e-05
					7	2.48139676842e-05
					8	2.53270508256e-05
					9	2.45934966693e-05
					10	2.63224969136e-05
					11	2.65883469673e-05
					12	2.61120344541e-05
					13	2.68892748458e-05
					14	2.66861428305e-05
					15	2.7144143333e-05
					16	2.68208824836e-05
					17	2.7068153533e-05
					18	2.71689202457e-05
					19	2.80819817421e-05
					20	2.83468062578e-05
				};
				\draw[hl]foreach\i in{10,20,...,80}{(0,0)--(\i:10cm)};
			\end{axis}
		}
		\caption{
			We write a python script to support Theorem~\ref{thm:main}.
			The script:
			1) set $I(W)=0.618$;
			2) loop for $n=1,\dotsc,20$;
			3) for each $n$, calculates $ε$ by $2^{-n/5}$; and
			4) profiles the process of $2^{20}$ uses of $W$.
			(That is, for block length $2^n$, it tests $2^{20-n}$ blocks.)
			For each $n$, the total times the script takes
			is divided by the dimension of the code and plotted above.
			The gray thin lines shows
			the expected behavior of classical polar codes---%
			their per-bit time should be $O(\log n)=O(n)$, i.e., rays.
			Our codes, however, do not follow any of gray thin lines
			but bends downward and crosses several gray thin lines.
			This matches the claimed $O(\log\log N)=O(\log n)$ behavior.
		}
		\label{fig:emptime}
	\end{figure}
	
	For the close-to-$1$ part of 2), \cite[Lemma~3.1]{BGS18} shows that
	$ℙ｛I_n≤2^{-2^{βn}}｝≥I(W)-O(2^{-n/μ})$ for some constants $β,μ$.
	One can also prove it barehanded using the same trick
	we used for the small-$I_i$ part of 2).
	This is the last piece of the proof.
	Now 2) is finished.
	The proof completes.

\subsection{Simulation}
	
	\begin{figure}
		\tikz{
			\begin{axis}[simu,
				         ylabel={$𝔼[τ]$},
				         xlabel={$n=\log_2(\text{block length})$}]
				\addplot table{
					0	0.0
					1	0.0
					2	1.5
					3	2.75
					4	3.375
					5	4.5625
					6	5.125
					7	5.828125
					8	6.5234375
					9	6.96875
					10	7.90625
					11	8.3544921875
					12	8.7314453125
					13	9.26879882812
					14	9.63500976562
					15	9.91046142578
					16	10.3467407227
					17	10.6794891357
					18	10.9180297852
					19	11.3841133118
					20	11.5504550934
					21	11.7551832199
					22	11.9149127007
					23	12.0850632191
					24	12.255657196
					25	12.4046368599
				};
				\draw[hl]foreach\i in{10,20,...,80}{(0,0)--(\i:10cm)};
			\end{axis}
		}
		\caption{
			We write a python script that
			grows the tree using Framed Rule~(\ref{eq:rule})
			and compute the exact $𝔼[τ]$ accordingly.
			We believe $𝔼[τ]$ is a good substitution of $\bC$
			by the reasoning before Formula~(\ref{eq:Etau}).
			Notice that the plot of $𝔼[τ]$
			bends downward like $\log n$ does.
			For one: it deviates from the expected behavior of
			classical polar codes (gray thin lines; linear behavior).
			For two: this reassures $𝔼[τ]=O(\log\log N)=O(\log n)$.
		}
		\label{fig:exptau}
	\end{figure}
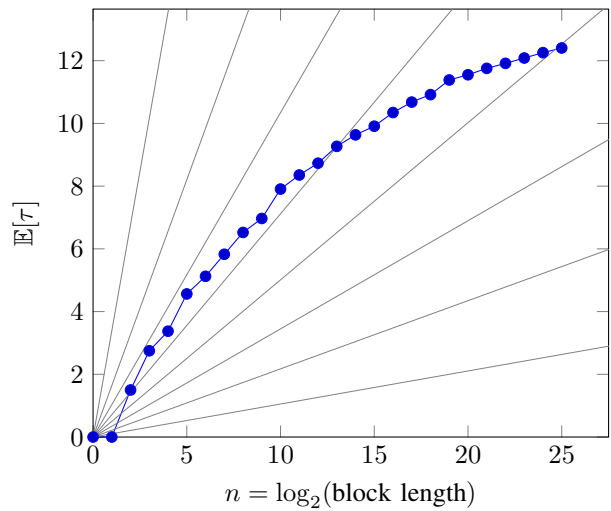
	
	\begin{figure}
		\tikz{
			\begin{axis}[simu,
				         ylabel={Sample mean of $τ$},
				         xlabel={$n=\log_2(\text{block length})$}]
				\addplot table{
					0	0.0
					1	0.0
					2	1.489
					3	2.723
					4	3.341
					5	4.563
					6	5.132
					7	5.854
					8	6.506
					9	6.89
					10	7.84
					11	8.324
					12	8.712
					13	9.405
					14	9.849
					15	9.869
					16	10.391
					17	10.791
					18	10.926
					19	11.366
					20	11.52
					21	11.92
					22	12.171
					23	12.033
					24	12.211
					25	12.123
					26	12.55
					27	12.59
					28	12.564
					29	13.112
					30	13.166
					31	13.31
					32	13.227
					33	13.387
					34	13.228
					35	13.303
					36	13.465
					37	13.751
					38	14.383
					39	13.857
					40	14.158
					41	14.461
					42	14.412
					43	14.158
					44	14.367
					45	14.165
					46	14.603
					47	14.714
					48	14.654
					49	14.572
					50	14.422
					51	14.565
					52	14.578
					53	14.834
					54	14.501
					55	14.81
					56	14.611
					57	14.84
					58	14.767
					59	14.472
					60	14.844
					61	14.876
					62	14.986
					63	14.927
				};
				\draw[hl]foreach\i in{10,20,...,80}{(0,0)--(\i:10cm)};
			\end{axis}
		}
		\caption{
			We write a python script that
			samples the process $Z_{i∧τ}$ a thousand times.
			The empirical $τ$ is plotted above.
			The result shows a clear trend
			that looks like $O(\log\log N)=O(\log n)$.
			This is again what we expected.
		}
		\label{fig:emptau}
	\end{figure}
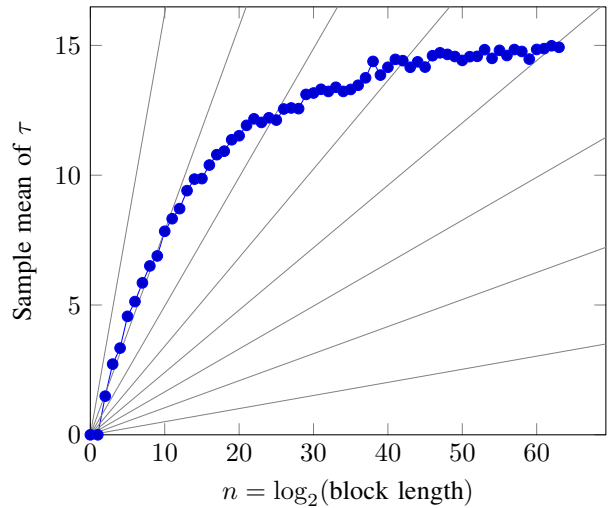
	
	We write a python script to support Theorem~\ref{thm:main}.
	The script:
	1)	sets $I(W)=0.618$;
	2)	loops for $n=1,\dotsc,20$;
	3)	for each $n$, evaluate $2^{-n/5}$ as $ε$ ; and
	4)	profiles the process of $2^{20}$ uses of $W$.
	(That is, for block length $2^n$, it tests $2^{20-n}$ blocks.)
	The empirical $\bC$ is shown in Fig.~\ref{fig:emptime}.
	Notice that the plot of $\bC$ \emph{does not} grow proportionally to $n$.
	For classical polar codes, in contrast,
	$\bC$ is proportional to $O(\log N)=O(n)$.
	
	For $n>20$, we do not test actual codes.
	Instead, we grow the tree using Framed Rule~(\ref{eq:rule})
	and compute $𝔼[τ]$ accordingly.
	We believe $𝔼[τ]$ is a good substitution of $\bC$
	by the reasoning before Formula~(\ref{eq:Etau}).
	Notice that the plot of $𝔼[τ]$ bends downward
	like $\log\log N≈\log n$ does.
	For one: it deviates from the expected behavior of classical polar codes.
	For two: this reassures $𝔼[τ]=O(\log\log N)=O(\log n)$
	as shown in Theorem~\ref{thm:grow}.
	At $n=25$, our construction reduces
	the number of butterfly devices by one-half.
	
	Starting from $n=26$ it is difficult
	to calculate the exact value of $𝔼[τ]$.
	We instead sample the process $Z_{i∧τ}$ a thousand times
	and accumulate the sample mean of $τ$.
	The result shows a clear trend that looks like $O(\log\log N)=O(\log n)$,
	which is what we expected.

\bibliographystyle{IEEEtran}
\IEEEtriggeratref{14}
\bibliography{LoglogTime-7}

\begin{thebibliography}{10}
\providecommand{\url}[1]{#1}
\csname url@samestyle\endcsname
\providecommand{\newblock}{\relax}
\providecommand{\bibinfo}[2]{#2}
\providecommand{\BIBentrySTDinterwordspacing}{\spaceskip=0pt\relax}
\providecommand{\BIBentryALTinterwordstretchfactor}{4}
\providecommand{\BIBentryALTinterwordspacing}{\spaceskip=\fontdimen2\font plus
\BIBentryALTinterwordstretchfactor\fontdimen3\font minus
  \fontdimen4\font\relax}
\providecommand{\BIBforeignlanguage}[2]{{%
\expandafter\ifx\csname l@#1\endcsname\relax
\typeout{** WARNING: IEEEtran.bst: No hyphenation pattern has been}%
\typeout{** loaded for the language `#1'. Using the pattern for}%
\typeout{** the default language instead.}%
\else
\language=\csname l@#1\endcsname
\fi
#2}}
\providecommand{\BIBdecl}{\relax}
\BIBdecl

\bibitem{AW10}
Y.~Altu{\u g} and A.~B. Wagner, ``Moderate deviation analysis of channel
  coding: Discrete memoryless case,'' in \emph{2010 IEEE International
  Symposium on Information Theory}, June 2010, pp. 265--269.

\bibitem{PV10}
Y.~Polyanskiy and S.~Verd{\'u}, ``Channel dispersion and moderate deviations
  limits for memoryless channels,'' in \emph{2010 48th Annual Allerton
  Conference on Communication, Control, and Computing (Allerton)}, Sept 2010,
  pp. 1334--1339.

\bibitem{AW14}
Y.~Altu{\u g} and A.~B. Wagner, ``Moderate deviations in channel coding,''
  \emph{IEEE Transactions on Information Theory}, vol.~60, no.~8, pp.
  4417--4426, Aug 2014.

\bibitem{Arikan15}
E.~Arikan, ``A packing lemma for polar codes,'' in \emph{2015 IEEE
  International Symposium on Information Theory (ISIT)}, June 2015, pp.
  2441--2445.

\bibitem{HT15}
M.~Hayashi and V.~Y.~F. Tan, ``Erasure and undetected error probabilities in
  the moderate deviations regime,'' in \emph{2015 IEEE International Symposium
  on Information Theory (ISIT)}, June 2015, pp. 1821--1825.

\bibitem{KKMPSU17}
S.~{Kudekar}, S.~{Kumar}, M.~{Mondelli}, H.~D. {Pfister}, E.~{{\c S}a{\c s}o{\v
  g}lu}, and R.~L. {Urbanke}, ``Reed--muller codes achieve capacity on erasure
  channels,'' \emph{IEEE Transactions on Information Theory}, vol.~63, no.~7,
  pp. 4298--4316, July 2017.

\bibitem{KRU13}
S.~Kudekar, T.~Richardson, and R.~L. Urbanke, ``Spatially coupled ensembles
  universally achieve capacity under belief propagation,'' \emph{IEEE
  Transactions on Information Theory}, vol.~59, no.~12, pp. 7761--7813, Dec
  2013.

\bibitem{PSU05}
H.~D. {Pfister}, I.~{Sason}, and R.~{Urbanke}, ``Capacity-achieving ensembles
  for the binary erasure channel with bounded complexity,'' \emph{IEEE
  Transactions on Information Theory}, vol.~51, no.~7, pp. 2352--2379, July
  2005.

\bibitem{PS07}
H.~D. {Pfister} and I.~{Sason}, ``Accumulate-repeat-accumulate codes:
  Capacity-achieving ensembles of systematic codes for the erasure channel with
  bounded complexity,'' \emph{IEEE Transactions on Information Theory},
  vol.~53, no.~6, pp. 2088--2115, June 2007.

\bibitem{Arikan09}
E.~Arikan, ``Channel polarization: A method for constructing capacity-achieving
  codes for symmetric binary-input memoryless channels,'' \emph{IEEE
  Transactions on Information Theory}, vol.~55, no.~7, pp. 3051--3073, July
  2009.

\bibitem{GX13}
V.~Guruswami and P.~Xia, ``Polar codes: Speed of polarization and polynomial
  gap to capacity,'' in \emph{2013 IEEE 54th Annual Symposium on Foundations of
  Computer Science}, Oct 2013, pp. 310--319.

\bibitem{MHU16}
M.~Mondelli, S.~H. Hassani, and R.~L. Urbanke, ``Unified scaling of polar
  codes: Error exponent, scaling exponent, moderate deviations, and error
  floors,'' \emph{IEEE Transactions on Information Theory}, vol.~62, no.~12,
  pp. 6698--6712, Dec 2016.

\bibitem{FT17}
\BIBentryALTinterwordspacing
S.~L. Fong and V.~Y.~F. Tan, ``Scaling exponent and moderate deviations
  asymptotics of polar codes for the awgn channel,'' \emph{Entropy}, vol.~19,
  no.~7, 2017. [Online]. Available:
  \url{http://www.mdpi.com/1099-4300/19/7/364}
\BIBentrySTDinterwordspacing

\bibitem{WD18m}
\BIBentryALTinterwordspacing
H.~Wang and I.~Duursma, ``Polar code moderate deviation: Recovering the scaling
  exponent,'' \emph{CoRR}, vol. abs/1806.02405, 2018. [Online]. Available:
  \url{http://arxiv.org/abs/1806.02405}
\BIBentrySTDinterwordspacing

\bibitem{BGS18}
J.~{B{\l}asiok}, V.~{Guruswami}, and M.~{Sudan}, ``{Polar Codes with
  exponentially small error at finite block length},'' \emph{ArXiv e-prints},
  Oct. 2018.

\bibitem{WD18l}
\BIBentryALTinterwordspacing
H.~Wang and I.~Duursma, ``Polar-like codes and asymptotic tradeoff among block
  length, code rate, and error probability,'' \emph{CoRR}, vol. abs/1812.08112,
  2018. [Online]. Available: \url{http://arxiv.org/abs/1812.08112}
\BIBentrySTDinterwordspacing

\bibitem{AT09}
E.~Arikan and E.~Telatar, ``On the rate of channel polarization,'' in
  \emph{2009 IEEE International Symposium on Information Theory}, June 2009,
  pp. 1493--1495.

\bibitem{Gallager13}
R.~G. Gallager, \emph{Stochastic processes: theory for applications}.\hskip 1em
  plus 0.5em minus 0.4em\relax Cambridge University Press, 2013.

\bibitem{Durrett10}
\BIBentryALTinterwordspacing
R.~Durrett, \emph{Probability: Theory and Examples}, 4th~ed.\hskip 1em plus
  0.5em minus 0.4em\relax New York, NY, USA: Cambridge University Press, 2010.
  [Online]. Available: \url{https://services.math.duke.edu/~rtd/PTE/PTE4_1.pdf}
\BIBentrySTDinterwordspacing

\bibitem{FV14}
A.~Fazeli and A.~Vardy, ``On the scaling exponent of binary polarization
  kernels,'' in \emph{2014 52nd Annual Allerton Conference on Communication,
  Control, and Computing (Allerton)}, Sept 2014, pp. 797--804.

\bibitem{AYK11}
A.~Alamdar-Yazdi and F.~R. Kschischang, ``A simplified successive-cancellation
  decoder for polar codes,'' \emph{IEEE Communications Letters}, vol.~15,
  no.~12, pp. 1378--1380, December 2011.

\bibitem{ZZWZP15}
L.~Zhang, Z.~Zhang, X.~Wang, C.~Zhong, and L.~Ping, ``Simplified
  successive-cancellation decoding using information set reselection for polar
  codes with arbitrary blocklength,'' \emph{IET Communications}, vol.~9,
  no.~11, pp. 1380--1387, 2015.

\bibitem{ZZPYG14}
Y.~Zhang, Q.~Zhang, X.~Pan, Z.~Ye, and C.~Gong, ``A simplified belief
  propagation decoder for polar codes,'' in \emph{2014 IEEE International
  Wireless Symposium (IWS 2014)}, March 2014, pp. 1--4.

\bibitem{EKMFLK15}
M.~El-Khamy, H.~Mahdavifar, G.~Feygin, J.~Lee, and I.~Kang, ``Relaxed channel
  polarization for reduced complexity polar coding,'' in \emph{2015 IEEE
  Wireless Communications and Networking Conference (WCNC)}, March 2015, pp.
  207--212.

\bibitem{EKMFLK17}
------, ``Relaxed polar codes,'' \emph{IEEE Transactions on Information
  Theory}, vol.~63, no.~4, pp. 1986--2000, April 2017.

\bibitem{WLZZ15}
D.~Wu, A.~Liu, Q.~Zhang, and Y.~Zhang, ``Concatenated polar codes based on
  selective polarization,'' in \emph{2015 12th International Computer
  Conference on Wavelet Active Media Technology and Information Processing
  (ICCWAMTIP)}, Dec 2015, pp. 436--442.

\bibitem{EECB17}
A.~Elkelesh, M.~Ebada, S.~Cammerer, and S.~t.~Brink, ``Flexible length polar
  codes through graph based augmentation,'' in \emph{SCC 2017; 11th
  International ITG Conference on Systems, Communications and Coding}, Feb
  2017, pp. 1--6.

\bibitem{WYY18}
X.~Wu, L.~Yang, and J.~Yuan, ``Information coupled polar codes,'' in \emph{2018
  IEEE International Symposium on Information Theory (ISIT)}, June 2018, pp.
  861--865.

\bibitem{WYXY18}
X.~Wu, L.~Yang, Y.~Xie, and J.~Yuan, ``Partially information coupled polar
  codes,'' \emph{IEEE Access}, pp. 1--1, 2018.

\bibitem{SG13}
G.~Sarkis and W.~J. Gross, ``Increasing the throughput of polar decoders,''
  \emph{IEEE Communications Letters}, vol.~17, no.~4, pp. 725--728, April 2013.

\bibitem{SGVTG14}
G.~Sarkis, P.~Giard, A.~Vardy, C.~Thibeault, and W.~J. Gross, ``Fast polar
  decoders: Algorithm and implementation,'' \emph{IEEE Journal on Selected
  Areas in Communications}, vol.~32, no.~5, pp. 946--957, May 2014.

\bibitem{Arikan11}
E.~Arikan, ``Systematic polar coding,'' \emph{IEEE Communications Letters},
  vol.~15, no.~8, pp. 860--862, August 2011.

\bibitem{MT14}
R.~Mori and T.~Tanaka, ``Source and channel polarization over finite fields and
  reed-solomon matrices,'' \emph{IEEE Transactions on Information Theory},
  vol.~60, no.~5, pp. 2720--2736, May 2014.

\bibitem{BGNRS18}
\BIBentryALTinterwordspacing
J.~Blasiok, V.~Guruswami, P.~Nakkiran, A.~Rudra, and M.~Sudan, ``General strong
  polarization,'' \emph{CoRR}, vol. abs/1802.02718, 2018. [Online]. Available:
  \url{http://arxiv.org/abs/1802.02718}
\BIBentrySTDinterwordspacing

\bibitem{FM94}
M.~{Feder} and N.~{Merhav}, ``Relations between entropy and error
  probability,'' \emph{IEEE Transactions on Information Theory}, vol.~40,
  no.~1, pp. 259--266, Jan 1994.

\end{thebibliography}

\end{document}